\documentclass[journal=jctcce,manuscript=article]{achemso}

 \usepackage{graphicx}
 \usepackage{amssymb}
 \usepackage{amsmath}
 \usepackage{amsfonts}
 \usepackage{mathrsfs}
 \usepackage{dcolumn}
 \usepackage{physics}
 \usepackage{empheq}
 \usepackage[most]{tcolorbox}
 \usepackage{cancel}
 \usepackage{xcolor}
 \usepackage{natbib}
 \usepackage{mciteplus}
 \usepackage{multirow}
 \usepackage{subfig}
 \usepackage{lscape}
 \usepackage{dcolumn}
 \newcolumntype{d}{D{.}{.}{-1}}
 
 \newtcbox{\mymath}[1][]{%
 	nobeforeafter, math upper, tcbox raise base,
 	enhanced, colframe=blue!30!black,
 	colback=blue!30, boxrule=1pt,
 	#1}
 
\newcommand{\mbf}[1]{#1}

\newcommand{\beq}{\begin{equation}}
\newcommand{\eeq}{\end{equation}}

\title{Is Externally Corrected Coupled Cluster Always Better than the Underlying 
Truncated Configuration Interaction?}

\author{Ilias Magoulas}
\affiliation
{Department of Chemistry, Michigan State University, East Lansing, Michigan 48824, 
USA}

\author{Karthik Gururangan}
\affiliation
{Department of Chemistry, Michigan State University, East Lansing, Michigan 48824, 
USA}

\author{Piotr Piecuch}
\affiliation
{Department of Chemistry, Michigan State University, East Lansing, Michigan 48824, 
USA}
\alsoaffiliation
{Department of Physics and Astronomy, Michigan State University, East Lansing, 
Michigan 48824, USA}
\email{piecuch@chemistry.msu.edu}

\author{J. Emiliano Deustua}
\affiliation
{Department of Chemistry, Michigan State University, East Lansing, Michigan 48824,
USA}

\author{Jun Shen}
\affiliation
{Department of Chemistry, Michigan State University, East Lansing, Michigan 48824,
USA}

\date{\today}

\begin{document}
	
\begin{abstract}
The short answer to the question in the title is `no'. We identify classes of 
truncated configuration interaction (CI) wave functions for which the externally 
corrected coupled-cluster (ec-CC) approach using the three-body ($T_{3}$) and four-body ($T_{4}$) components 
of the cluster operator extracted from CI does not improve the results of the 
underlying CI calculations. Implications of our analysis, illustrated by numerical 
examples, for the ec-CC computations using truncated and selected CI methods are 
discussed. We also introduce a novel ec-CC approach
using the $T_{3}$ and $T_{4}$ amplitudes obtained with the selected CI
scheme abbreviated as CIPSI, correcting the resulting energies for
the missing $T_{3}$ correlations not captured by CIPSI with the help of
moment expansions similar to those employed in the completely renormalized CC
methods.
\end{abstract}

\section{INTRODUCTION}
\label{section1}

It is well-established that methods based on the exponential wave function ansatz
\cite{Hubbard1957a,Hugenholtz1957} of coupled-cluster (CC) theory,\cite{Coester:1958,cizek1,cizek2,cizek4}
\beq
|\Psi \rangle = e^{T} |\Phi \rangle ,
\label{eq-ccansatz}
\eeq
where
\beq
T = \sum_{n=1}^{N} T_{n}
\label{eq-clusterop}
\eeq
is the cluster operator, $T_{n}$ is the $n$-body component of $T$, $N$ is the number of correlated
electrons, and $|\Phi\rangle$ is the reference determinant defining the Fermi vacuum,
are among the most efficient ways of
incorporating many-electron correlation effects in molecular applications.\cite{paldus-li,Bartlett2007a}
However, the conventional and most practical single-reference CC methods, including
the CC singles and doubles (CCSD)
approach,\cite{ccsd,ccsd2} where $T$ is truncated at $T_{2}$, and the quasi-perturbative correction to CCSD due to
$T_{3}$ clusters defining the widely used CCSD(T) approximation,\cite{Raghavachari1989} fail in multi-reference
situations, such as bond breaking and strongly correlated systems (cf., e.g., refs
\citenum{paldus-li,Bartlett2007a,Piecuch1990}). In fact, no traditional truncation in the cluster operator
at a given many-body rank, including
higher-order CC methods, such as the CC approach with singles, doubles, and triples (CCSDT), where $T$ is truncated
at $T_{3}$,\cite{ccfullt,ccfullt2} and the CC approach with singles,
doubles, triples, and quadruples (CCSDTQ), where $T$ is truncated at $T_{4}$,\cite{ccsdtq0,ccsdtq2}
can handle systems with larger numbers of strongly correlated electrons.\cite{Podeszwa2002,Degroote2016}
Since conventional multi-reference 
methods
\cite{paldus-li,Bartlett2007a,Lyakh2012a,evangelista-perspective-jcp-2018}
may be inapplicable to such problems as well (in part due to rapidly growing dimensionalities of
the underlying active spaces), it is worth exploring
alternative
ideas, including those that combine different wave function ans{\" a}tze,
which would allow us to provide an accurate and balanced description of nondynamic and dynamic correlations
in a wide range of many-electron systems encountered in chemical applications.

One of the interesting ways of improving the results of single-reference CC calculations in multi-reference
and strongly correlated situations, which is based on combining the CC and non-CC (e.g., configuration
interaction (CI)) concepts and which is
the main topic of this study, is the externally corrected CC (ec-CC)
framework
\cite{Paldus1984a,Piecuch1996a,Paldus1994,Stolarczyk1994,Peris1997,Peris1999,Li1997,Li1998,%
LiPaldus2006,Deustua2018,Aroeira2020,chan2021ecCC}
(see ref \citenum{Paldus2017} for a review).
The ec-CC approaches are based on the observation that as long as the electronic Hamiltonian $H$ does
not contain higher--than--two-body interactions, the CC amplitude equations projected on the
singly and doubly excited determinants,
\beq
\mel{\Phi_i^a}{(H_Ne^T)_C}{\Phi} = 0
\label{cceq-singles}
\eeq
and
\beq
\mel{\Phi_{ij}^{ab}}{(H_Ne^T)_C}{\Phi} = 0,
\label{cceq-doubles}
\eeq
respectively, in which no approximations are made, do not engage the higher-rank $T_{n}$ components of the
cluster operator $T$ with $n > 4$. Thus, by solving these nonlinear, energy-independent, equations, which
can also be written as
\beq
\begin{split}
\bra{\Phi_i^a} \![&F_N + (F_N T_1)_C + (F_N T_2)_C + (F_N \tfrac{1}{2}
T_1^2)_C + (V_N T_1)_C + (V_N T_2)_C + (V_N \tfrac{1}{2} T_1^2)_C \\
& + (V_N T_3)_C + (V_N T_1 T_2)_C + (V_N \tfrac{1}{3!} T_1^3)_C]\! \ket{\Phi} = 0 ,
\end{split}
\label{fcc-singles}
\eeq
\beq
\begin{split}
\bra{\Phi_{ij}^{ab}} \![&(F_N T_2)_C + (F_N T_3)_C + (F_N T_1 T_2)_C
+ V_N + (V_N T_1)_C + (V_N T_2)_C + (V_N \tfrac{1}{2} T_1^2)_C \\
& + (V_N T_3)_C + (V_N T_1 T_2)_C + (V_N \tfrac{1}{3!} T_1^3)_C \\
& + (V_N T_4)_C + (V_N T_1 T_3)_C + (V_N \tfrac{1}{2} T_2^2)_C + (V_N
\tfrac{1}{2} T_1^2 T_2)_C + (V_N \tfrac{1}{4!} T_1^4)_C]\! \ket{\Phi} = 0 ,
\end{split}
\label{fcc-doubles}
\eeq
for the singly and doubly excited clusters, $T_{1}$ and $T_{2}$, respectively, in the presence
of their exact triply ($T_{3}$) and quadruply ($T_{4}$) excited counterparts extracted from
full CI (FCI), one obtains the exact $T_{1}$ and $T_{2}$ and the exact correlation energy
$\Delta E \equiv E - \langle \Phi | H | \Phi \rangle$,
which in the case of the Hamiltonians with two-body interactions is given by the expression
\beq
\Delta E = \mel{\Phi}{(H_Ne^T)_{FC}}{\Phi} = 
\mel{\Phi} {[(F_N T_1)_{FC} + (V_N T_2)_{FC} + (V_N \tfrac{1}{2} T_1^2)_{FC}]}{\Phi} .
\label{fcc-energy}
\eeq
This suggests that by using external wave functions capable of generating an accurate representation of
$T_{3}$ and $T_{4}$ clusters, and subsequently solving for $T_{1}$ and $T_{2}$
using eqs \ref{fcc-singles} and \ref{fcc-doubles},
one should not only produce correlation energies that are much better
than those obtained with CCSD, where $T_{3}$ and $T_{4}$ are zero, but also substantially
improve the results of
the calculations used to provide $T_{3}$ and $T_{4}$. 
Throughout this article,
we use the notation in which $|\Phi_{i_{1} \ldots i_{n}}^{a_{1} \ldots a_{n}} \rangle$
are the $n$-tuply excited determinants, with $i,j, \ldots$ and $a,b, \ldots$ representing the occupied
and unoccupied spin-orbitals in $|\Phi\rangle$, respectively,
$H_{N} = H - \langle \Phi | H | \Phi \rangle = F_{N} + V_{N}$
is the Hamiltonian in the normal-ordered form, with $F_{N}$ and $V_{N}$ representing its one- and two-body
components, and $(AB)_{FC}$, $(AB)_C$, and $AB$ are the fully connected, connected but
not fully connected, and disconnected products of operators $A$ and $B$, respectively.
In the language of diagrams, $(AB)_{FC}$ is a connected operator product having no external fermion lines,
$(AB)_C$ is a connected operator product with some fermion lines left uncontracted, and $AB$ implies that
no fermion lines connect $A$ and $B$.

As already alluded to above, one of the main premises of the ec-CC methodology is the expectation that as long as
the non-CC approach providing $T_{3}$ and $T_{4}$ clusters is not FCI, the ec-CC calculations improve the results
obtained with it.
One of the
two
objectives of this study is to examine the validity of
this premise.
By performing the appropriate mathematical analysis, backed by numerical examples, we
demonstrate
that, in addition to the exact, FCI or full CC, states, there
exist large classes of truncated CI and CC wave functions
which, after extracting $T_{1}$ through $T_{4}$ from them via the cluster analysis
procedure\cite{cizek-paldus-sroubkova-1969}
adopted in all ec-CC considerations, satisfy eqs \ref{fcc-singles} and \ref{fcc-doubles}.
In all those cases, which are further elaborated on in Section \ref{section2}, the ec-CC calculations return
back the energies obtained in the calculations that provide $T_{3}$ and $T_{4}$ clusters,
i.e., they offer no benefits whatsoever.
While it is obvious that any CC state with $T = \sum_{n=1}^{M} T_{n}$, where $2 \leq M \leq N$, including
the conventional CCSD ($M=2$), CCSDT ($M=3$), CCSDTQ ($M=4$), etc. truncations and their active-space CCSDt, CCSDtq, etc.
analogs,\cite{piecuch-qtp} which all treat the $T_{1}$ and $T_{2}$ components of $T$ fully,
satisfies eqs \ref{fcc-singles} and \ref{fcc-doubles}, the finding that there are truncated
CI states that result in the $T_{n}$ operators with $n = 1\mbox{--}4$ which are solutions of eqs
\ref{fcc-singles} and \ref{fcc-doubles} is
far less trivial. To prove it, one has to pay attention to various subtleties, such as the fact that
connected operator products of the $(AB)_{FC}$ and $(AB)_{C}$ types entering the ec-CC considerations via
eqs \ref{fcc-singles}--\ref{fcc-energy} are
not necessarily
connected
in a diagrammatic sense of the many-body perturbation theory
(MBPT), since operators $A$ and $B$ involved in forming $(AB)_{FC}$ or $(AB)_{C}$ may themselves be disconnected
or, even, unlinked. For example, the $T_{n}$ components of the cluster operator $T$, which enter
eqs \ref{fcc-singles}--\ref{fcc-energy}, are connected in a sense of MBPT diagrammatics if they
originate from standard CC computations or cluster analysis of FCI, but they may no longer
be connected if obtained from cluster analysis of truncated CI wave
functions. The
connected operator product only means that operators $A$ and $B$, each treated as a whole,
are connected with at least one fermion line, but the connectedness of
$(AB)_{FC}$ or
$(AB)_{C}$ in the MBPT sense depends on the contents of $A$ and $B$.
When using truncated CI wave functions as sources of $T_{3}$ and $T_{4}$ clusters,
as is the case in the present work, details of this kind become essential.

The formal and numerical results reported in this study may have
significant
implications for the development of future methods based on the ec-CC ideas.
The external sources of $T_{3}$ and $T_{4}$ clusters adopted in the
ec-CC methods developed to date include projected unrestricted Hartree--Fock wave functions, which
were used in the past to rationalize diagram cancellations defining the approximate coupled-pair approaches,
when applied to certain classes of strongly correlated model systems
\cite{Paldus1984a,Piecuch1991,Piecuch1996a} (see, also,
ref \citenum{Paldus2017} and references therein), and wave functions obtained with methods designed
to capture nondynamic correlation effects relevant to molecular applications, such as bond breaking
and polyradical species, including valence-bond,\cite{Paldus1994}
complete-active-space self-consistent-field (CASSCF),\cite{Stolarczyk1994,Peris1997}
multi-reference CI (MRCI),\cite{Li1997,Li1998,LiPaldus2006} perturbatively selected CI (PSCI),\cite{Peris1999}
FCI quantum Monte Carlo (FCIQMC),\cite{Deustua2018,Eriksen2020} adaptive CI (ACI),\cite{Aroeira2020}
density-matrix renormalization group (DMRG),\cite{chan2021ecCC} and heat-bath CI (HCI)\cite{chan2021ecCC}
approaches. One could also develop extensions of the ec-CC formalism by considering projections of the CC equations
on higher--than--doubly excited determinants and extracting the relevant $T_{n}$ components with $n \geq 4$
from a non-CC source, as in the ecCCSDt-CASSCF scheme of ref \citenum{ec-ccsdt-cas}, where the authors
corrected the CCSDt equations by the $T_{4}$- and $T_{5}$-containing terms extracted from CASSCF.
While some of the above ec-CC methods, especially the reduced multi-reference CCSD (RMRCCSD) approach,
\cite{Li1997,Li1998} which uses $T_{3}$ and $T_{4}$ clusters extracted from MRCI, and its RMRCCSD(T) extension
correcting the RMRCCSD energies for certain types of $T_{3}$ correlations missing in MRCI wave functions,
\cite{LiPaldus2006} the PSCI-based ec-CC scheme introduced in ref \citenum{Peris1999}, and the
cluster-analysis-driven FCIQMC (CAD-FCIQMC) approach of refs \citenum{Deustua2018,Eriksen2020}, which utilizes
$T_{3}$ and $T_{4}$ amplitudes extracted from stochastic wave function propagations defining the FCIQMC
framework,\cite{Booth2009,Cleland2010,Ghanem2019,ghanem_alavi_fciqmc_2020,ghanem_alavi_fciqmc_2021}
offer considerable improvements compared to both CCSD and the underlying
CI calculations providing $T_{3}$ and $T_{4}$ clusters (cf., e.g., refs
\citenum{Li1997,Li1998,LiPaldus2006,rmrcc3,rmrcc5,rmrcc7,rmrcc12,rmrcc-bnc2,rmrcc-biradicals} for
illustrative examples of successful RMRCC computations), there are situations where the
improvements are minimal or none.

The most recent example demonstrating
that the ec-CC computations do not necessarily outperform the underlying CI calculations is the
ACI-CC method implemented in ref \citenum{Aroeira2020},
in which $T_{3}$ and $T_{4}$ clusters are
extracted from the wave functions obtained with the ACI approach.\cite{adaptive_ci_1,adaptive_ci_2}
As shown, for example, in Table 2 of ref \citenum{Aroeira2020}, the ACI-CCSD calculations worsen the
underlying ACI results for the automerization of cyclobutadiene so much that there is virtually no
difference between the ACI-CCSD and poor CCSD energetics (see, also, Figure 5 in ref \citenum{Aroeira2020}).
The enrichment of the ACI wave functions using MRCI-like arguments through the
extended ACI approach abbreviated by the authors of ref \citenum{Aroeira2020} as xACI,
followed by cluster analysis
to obtain $T_{3}$ and $T_{4}$ and ec-CC iterations to determine $T_{1}$ and $T_{2}$,
defining the xACI-CCSD method, improves the ACI-CCSD barrier heights, but the xACI-CCSD
computations do not improve the corresponding ACI and xACI results. Similar remarks apply
to the potential energy curve and vibrational term values of the beryllium dimer, shown
in Figure 2 and Table 1 of ref \citenum{Aroeira2020}, where there is virtually no
difference between the CCSD(T) and ACI-CCSD(T) data (ACI-CCSD(T) stands for the ACI-CCSD
calculations corrected for the $T_{3}$ correlations missing in the ACI wave functions).
This shows that there are classes of truncated CI wave functions that are either good enough
in their own right or that have a specific mathematical structure such that the ec-CC calculations
using them do not offer any significant benefits, while adding to the computational costs.

This leads us to the second objective of the present study, discussed mostly in Section \ref{section3},
namely, the exploration of an alternative to ACI,
abbreviated as CIPSI, which stands for the CI method using perturbative selection made iteratively,
\cite{sci_3,cipsi_1,cipsi_2} as a source of $T_{3}$ and $T_{4}$ clusters in the ec-CC considerations.
In analogy to
the ACI approach explored in the ec-CC context in ref \citenum{Aroeira2020},
the HCI method utilized in the ec-CC calculations reported in ref \citenum{chan2021ecCC},
and the PSCI scheme adopted in the ec-CC study discussed in ref \citenum{Peris1999},
CIPSI belongs to the broader category of selected CI approaches, which date back to
the late 1960s and early 1970s\cite{sci_1,sci_2,sci_3,sci_4} and which have recently attracted renewed
attention.\cite{adaptive_ci_1,adaptive_ci_2,asci_1,asci_2,ici_1,ici_2,shci_1,shci_2,shci_3,cipsi_1,cipsi_2}
In addition to using CIPSI to support parts of our mathematical analysis
in Section \ref{section2},
we demonstrate that the ec-CC approach using $T_{3}$ and $T_{4}$ cluster components
extracted from the CIPSI wave functions improves the underlying CIPSI results,
especially after correcting the ec-CC energies for the missing $T_{3}$ correlations. On the other hand,
as shown
in Section \ref{section3},
the CIPSI energies
corrected using multi-reference second-order MBPT can be competitive with the
corresponding CIPSI-driven ec-CC computations, at least when smaller molecular
systems are examined. This agrees with the excellent performance of the perturbatively
corrected CIPSI approach, when compared with other methods aimed at near-FCI
energetics, observed,
for example, in ref \citenum{cipsi_benzene},
which reinforces the importance of the question posed in the
title.

\section{CLASSES OF CI WAVE FUNCTIONS THAT SATISFY \lowercase{ec}-CC EQUATIONS}
\label{section2}

We begin
with the formal considerations,
supported by the numerical evidence shown in
Tables \ref{table1}--\ref{table3}, aimed at identifying the non-exact ground-state wave functions $|\Psi\rangle$
that, after performing cluster analysis on them, result in the $T_{1}$ through $T_{4}$ components satisfying
eqs \ref{fcc-singles} and \ref{fcc-doubles} and returning back the energies associated with these
$|\Psi\rangle$ states. The mathematics of the ec-CC framework, focusing on the ec-CC approaches
using truncated CI wave functions to generate $T_{3}$ and $T_{4}$ clusters, and its implications
are discussed in Section \ref{section2.1}. The calculations illustrating the ec-CC theory aspects
examined in Section \ref{section2.1} are discussed in Section \ref{section2.2}.

\subsection{Mathematical Analysis of the \lowercase{ec}-CC Formalism}
\label{section2.1}

We have already noted that any state resulting from
the CC calculations using $T = \sum_{n=1}^{M} T_{n}$, where $2\leq M \leq N$, starting with the basic CCSD
approximation and including the remaining members of the conventional CCSD, CCSDT, CCSDTQ, etc. hierarchy,
satisfies eqs \ref{fcc-singles} and \ref{fcc-doubles}. In fact, any wave function $|\Psi\rangle$ that
uses the exponential CC ansatz defined by eq \ref{eq-ccansatz} and treats $T_{1}$ and $T_{2}$ clusters
fully satisfies these equations too. This alone, while obvious to CC practitioners, might already be
a potential issue in the context of ec-CC considerations, since, at least in principle, one can envision
situations where some non-exact, non-CC approaches recreate, to a good approximation, such CC states
and energies associated with them, diminishing the value of the corresponding ec-CC calculations.

As elaborated on in this subsection, and as demonstrated in Appendices A and B which contain the relevant
mathematical proofs, similar statements apply to certain classes of truncated CI approaches, when used
as providers of $T_{3}$ and $T_{4}$ clusters for ec-CC computations. In particular, if we use any
conventional CI truncation to define the ground state $|\Psi\rangle$ in which singly and doubly excited
contributions are treated fully, as in the CISD, CISDT, CISDTQ, etc. approaches, i.e.,
\beq
|\Psi \rangle = (1 + C) |\Phi \rangle ,
\label{eq-ciansatz}
\eeq
where we assumed the intermediate normalization and where
\beq
C = \sum_{n=1}^{M} C_{n}
\label{ci-operator}
\eeq
is the corresponding excitation operator, with $C_{n}$ representing its $n$-body component and $2 \leq M \leq N$,
and then, after determining the CI excitation amplitudes by diagonalizing the Hamiltonian,
define the cluster operator $T$ as
\beq
T = \ln (1 + C) = \sum_{m=1}^{N} \tfrac{(-1)^{m-1}}{m} C^m ,
\label{t-definition}
\eeq
to bring the CI expansion, eq \ref{eq-ciansatz}, to an exponential form, eq \ref{eq-ccansatz}, which leads to the
well-known\cite{cizek1,cizek4,cizek-paldus-sroubkova-1969}
definitions of the $T_{1}$ through $T_{4}$ components adopted in all ec-CC methods,
\beq
  \begin{split}
  T_1 &= C_1, \\
  T_2 &= C_2 - \tfrac{1}{2} C_1^2, \\
  T_3 &= C_3 - C_1 C_2 + \tfrac{1}{3} C_1^3, \\
  T_4 &= C_4 - C_1 C_3 - \tfrac{1}{2} C_2^2 + C_1^2 C_2 - \tfrac{1}{4} C_1^4 ,
\end{split}
\label{cluster_analysis}
\eeq
the resulting $T_{n}$, $n = 1\mbox{--}4$, amplitudes satisfy eqs \ref{fcc-singles} and \ref{fcc-doubles}.
In other words, if we extract the $T_{3}$ and $T_{4}$ components of $T$ through the cluster analysis
of the CI state $|\Psi\rangle$ defined by eqs \ref{eq-ciansatz} and \ref{ci-operator}, using the
relationships between the $C_{n}$ and $T_{n}$ operators given by eq \ref{cluster_analysis}, and solve
for $T_{1}$ and $T_{2}$ in the presence of $T_{3}$ and $T_{4}$ obtained in this way, we recover the
truncated CI energy associated with $|\Psi\rangle$ back, without improving it at all. This means that if
we follow the above recipe, without making any additional {\it a posteriori} modifications in $T_{3}$ and $T_{4}$,
which we refer to as variant I of ec-CC, abbreviated throughout this paper as ec-CC-I, the ec-CC calculations using
the CISD, CISDT, CISDTQ, etc. wave functions to generate $T_{3}$ and $T_{4}$ with the help of eq \ref{cluster_analysis}
reproduce the corresponding CI energies, nothing more. At first glance, the ec-CC-I calculations
using the CISD and CISDT states as external sources of $T_{3}$ and $T_{4}$ amplitudes seem strange, but, as
further clarified by the proofs presented in Appendices A and B, there is nothing strange about it.
If we do not impose any constraints on the $T_{3}$ and $T_{4}$ components resulting from
the cluster analysis defined by eq \ref{cluster_analysis}, the ec-CC calculations using the CISD or CISDT
wave functions are as legitimate as all others. The unconstrained ec-CC-I calculations using the
CISD and CISDT wave functions in a cluster analysis return back the corresponding CISD and CISDT energies, since the
ec-CC-I scheme, as summarized above, allows for the purely disconnected $T_{3}$ and $T_{4}$ amplitudes, such as
\beq
T_3 = - C_1 C_2 + \tfrac{1}{3} C_1^3
\label{t3-DC}
\eeq
or
\beq
T_4 = - C_1 C_3 - \tfrac{1}{2} C_2^2 + C_1^2 C_2 - \tfrac{1}{4} C_1^4 .
\label{t4-DC}
\eeq
This undesirable feature of the ec-CC-I scheme is a consequence of artificially imposing the exponential structure of
the wave function on a truncated CI expansion, which does not have it. The only conventional CI
state that can be represented
by the connected $T_{n}$ components by exploiting the relations between the $C_{n}$ and $T_{n}$ operators
given by eq \ref{cluster_analysis} is a FCI state.

In reality, the ec-CC-I scheme, as defined above, is never used in the context of ec-CC calculations based
on $T_{3}$ and $T_{4}$ extracted from truncated CI, since allowing the purely disconnected forms of
$T_{3}$ and $T_{4}$ operators, when the corresponding $C_{3}$ and $C_{4}$ amplitudes are zero,
as in eqs \ref{t3-DC} and \ref{t4-DC}, is
problematic (we recall that in the exact, FCI, description, all many-body components
of the cluster operator $T$ are connected\cite{Hubbard1957a,Hugenholtz1957}). 
To eliminate the risk of introducing the purely disconnected three-
and four-body components of the cluster operator $T$ into the ec-CC equations for $T_{1}$ and $T_{2}$, eqs
\ref{fcc-singles} and \ref{fcc-doubles}, in all practical implementations of the ec-CC methodology employing
truncated CI wave functions, such as those reported in refs
\citenum{Peris1999,Li1997,Li1998,LiPaldus2006,Aroeira2020,chan2021ecCC},
one keeps only those $T_{3}$ and $T_{4}$
amplitudes resulting from the cluster analysis for which the corresponding $C_{3}$ and $C_{4}$
excitation coefficients are nonzero.\cite{Li1997} This does not guarantee a complete elimination of
disconnected diagrams from the resulting $T_{3}$ and $T_{4}$ amplitudes, since all Hamiltonian
diagonalizations using conventional CI truncations result in unlinked wave function contributions that
do not cancel out, but it does take care of the negative consequences associated with the presence of the
purely disconnected $T_{3}$ and $T_{4}$ terms that fall into the category of expressions represented by eqs
\ref{t3-DC} and \ref{t4-DC}. The resulting ec-CC protocol, in which one does not allow the problematic 
$T_{3}$ and $T_{4}$ components that do not have the companion $C_{3}$ and $C_{4}$ amplitudes, is called
in this paper variant II of ec-CC, abbreviated as ec-CC-II. The RMRCCSD approach introduced in
ref \citenum{Li1997}, the ACI-CCSD scheme implemented in ref \citenum{Aroeira2020}, and the CIPSI-driven ec-CC-II
algorithm discussed in Section \ref{section3} are the examples of ec-CC methods in this category. The removal of
certain classes of triples and quadruples from the ec-CC considerations, which results from imposing the
above constraint on the $T_{3}$ and $T_{4}$ amplitudes allowed in eqs \ref{fcc-singles} and \ref{fcc-doubles},
can be compensated by correcting the ec-CC-II energies for the missing $T_{3}$ or $T_{3}$ and $T_{4}$ correlations,
as in the CCSD(T)-like triples corrections to RMRCCSD and ACI-CCSD adopted in refs \citenum{LiPaldus2006} and
\citenum{Aroeira2020}, defining the RMRCCSD(T) and ACI-CCSD(T) approaches, respectively, and the CIPSI-based
ec-CC-II$_3$ method introduced in Section \ref{section3}, which relies on the more accurate
and more robust moment corrections
(cf., also, ref \citenum{chan2021ecCC}).
We will return to the issue of correcting the ec-CC-II energies for the missing $T_{3}$ correlations
when discussing our new CIPSI-driven ec-CC-II approach in the next section.
To facilitate reading of the rest of this article,
the three categories of the ec-CC models analyzed in this work are summarized in Table \ref{table4}.

The ec-CC-II protocol eliminates problems resulting from allowing the purely disconnected forms of $T_{3}$ and $T_{4}$
operators in the ec-CC calculations, when the corresponding $C_{3}$ and $C_{4}$ amplitudes are zero,
but it does not prevent the collapse of the ec-CC energies onto their CI counterparts. The ec-CC-II
computations, in which the $T_{3}$ and $T_{4}$ clusters entering eqs \ref{fcc-singles} and \ref{fcc-doubles}
are extracted from CI calculations using a complete treatment
of singles, doubles, triples, and quadruples, as in the CISDTQ, CISDTQP, CISDTQPH, etc. truncations,
where $M$ in eq \ref{ci-operator} is at least 4 and letters `P' and `H'
in the CISDTQP and CISDTQPH acronyms stand for pentuples ($C_{5}$) and hextuples ($C_{6}$), respectively,
return back the corresponding CI energies. In all of these cases, the ec-CC methodology, including
even its most proper ec-CC-II variant, offers no benefits whatsoever, while adding to the computational costs
associated with cluster analysis of the underlying CI wave functions
and dealing with eqs \ref{fcc-singles} and \ref{fcc-doubles} after
the respective CI Hamiltonian diagonalizations.
This is because once all singles, doubles, triples, and quadruples are included in CI calculations,
every nonzero $T_{n}$, $n = 3,4$, amplitude resulting from the cluster analysis of the underlying
CI state has a companion, also nonzero, $C_{n}$ excitation coefficient, i.e., the ec-CC-I and ec-CC-II
schemes become equivalent. One might argue that the above observation does not diminish the usefulness of
ec-CC computations, since one never uses high-level single-reference CI methods, such as CISDTQ, as
sources of $T_{3}$ and $T_{4}$ amplitudes in practical applications, which is a correct statement,
but a remark like that could be misleading.
Nowadays, one can generate approximate CI-type wave functions that provide a highly accurate
representation of the $C_{n}$ components through quadruples and beyond in computationally efficient
ways via stochastic CIQMC
propagations\cite{Booth2009,Cleland2010,Ghanem2019,ghanem_alavi_fciqmc_2020,ghanem_alavi_fciqmc_2021}
and semi-stochastic implementations of selected CI techniques, as in
HCI
\cite{shci_1,shci_2,shci_3}
and
the modern formulation
of CIPSI.\cite{cipsi_1,cipsi_2}
As shown in Section \ref{section3},
the ec-CC-II$_3$ approach using the $T_{3}$ and $T_{4}$ clusters extracted from the CIPSI wave functions
and corrected for the missing $T_{3}$ correlations does not have to improve the corresponding CIPSI
energetics corrected using second-order MBPT when smaller many-electron problems are examined.

If we limit ourselves to conventional CI truncations defined by eqs
\ref{eq-ciansatz} and \ref{ci-operator}, the only situations where the ec-CC-I and ec-CC-II
protocols differ, giving the ec-CC-II approach a chance to improve the results of the corresponding
ec-CC-I and CI computations, are those in which $T_{3}$ and $T_{4}$ clusters are extracted from CISD or
CISDT. To be more specific, when one uses the CISD wave function, where $C_{3}$ and $C_{4}$ are by definition zero,
in the ec-CC-II calculations, which means that $T_{3}$ and $T_{4}$ in eqs \ref{fcc-singles} and \ref{fcc-doubles}
are set at zero as well, the ec-CC-II energy becomes equivalent to that obtained with CCSD,
as opposed to the usually less accurate CISD value obtained with ec-CC-I. When the CISDT wave function
is employed in the ec-CC-II computations, one solves eqs \ref{fcc-singles} and \ref{fcc-doubles} for
$T_{1}$ and $T_{2}$ in the presence of $T_{3}$ contributions having companion $C_{3}$ excitation amplitudes
extracted from
CISDT. This
may lead to major improvements compared to the corresponding ec-CC-I calculations
that return back the CISDT energy and the results obtained with CCSD, in which $T_{3}$ is
ignored. Unfortunately,
since $C_{4} = 0$ in CISDT, i.e., $T_{4}$ correlations are not accounted for, the CISDT-based ec-CC-II
approach may produce erratic results in more complex multi-reference situations. We will return to
the discussion of these accuracy patterns, including the equivalence of the ec-CC-I, ec-CC-II, and the
underlying CI approaches when the ec-CC calculations use wave functions characterized by a full treatment
of $C_{n}$ amplitudes with $n = 1\mbox{--}4$,
in Section \ref{section2.2}.

As shown in Appendices A and B, the above relationships between the results of the ec-CC calculations
using $T_{3}$ and $T_{4}$ clusters extracted from truncated CI computations and the corresponding CI energetics
can be generalized to unconventional truncations in the linear excitation operator $C$ defining the wave
function $|\Psi\rangle$ via eq \ref{eq-ciansatz} as long as the $C_{1}$ and $C_{2}$ components of $C$ contain
a complete set of singles and doubles. To state this generalization, which is the heart of the mathematical
and numerical analysis presented in this section, more precisely, let us consider the CI eigenvalue
problem in which the ground electronic state is defined as follows:
\beq
\ket{\Psi} = (1 + C^{(P_\text{A})} + C^{(P_\text{B})}) \ket{\Phi},
\label{eq-genciansatz}
\eeq
where the singly and doubly excited components of $|\Psi\rangle$, described by the $C^{(P_\text{A})}$ operator,
are treated fully, i.e.,
\beq
C^{(P_\text{A})} = C_1 + C_2 ,
\label{eq-CPA}
\eeq
and where the remaining wave function contributions, if any, are represented by
\beq
C^{(P_\text{B})} = \sum_{n \geq 3} C_n^{(P_\text{B})}
\label{eq-CPB}
\eeq
(as in the case of eq \ref{eq-ciansatz}, we impose the intermediate normalization on $|\Psi\rangle$).
We do not make any specific assumptions regarding the $C^{(P_\text{B})}$ operator other than
the requirement that it does not contain the one- and two-body components of $C$, which are included in
$C^{(P_\text{A})}$. The above definitions encompass the previously discussed conventional CI
truncations, starting from CISD, where $C^{(P_\text{B})} = 0$, and including the remaining
members of the CISD, CISDT, CISDTQ, etc. hierarchy, for which the relevant $n$-body components
of $C^{(P_\text{B})}$ are treated fully, and the various selected CI approaches with all singles
and doubles and subsets of higher--than--double excitations. To facilitate our discussion below,
and to aid the presentation of the proofs in Appendices A and B, we designate the subspace of the
many-electron Hilbert space $\mathscr{H}$ spanned by the singly and doubly excited determinants,
$|\Phi_{i}^{a}\rangle$ and $|\Phi_{ij}^{ab}\rangle$, respectively, which are jointly abbreviated as
$\ket{\Phi_\alpha}$ and which match the content of the $C^{(P_\text{A})}$ operator defined by eq \ref{eq-CPA},
as $\mathscr{H}^{(P_\text{A})}$. The subspace of $\mathscr{H}$ spanned by the determinants
corresponding to the content of $C^{(P_\text{B})}$, eq \ref{eq-CPB}, denoted as $\ket{\Phi_\beta}$, is
designated as $\mathscr{H}^{(P_\text{B})}$, and the orthogonal complement to
${\mathscr H}^{(P)} \oplus \mathscr{H}^{(P_\text{A})} \oplus \mathscr{H}^{(P_\text{B})}$,
which contains the remaining determinants $\ket{\Phi_\gamma}$ not included in the CI wave function
$|\Psi\rangle$, is denoted as ${\mathscr H}^{(Q)}$ (${\mathscr H}^{(P)}$ is a one-dimensional subspace
of ${\mathscr H}$ spanned by the reference determinant $|\Phi\rangle$). Using the
above notation, we can write the CI eigenvalue problem for the ground-state wave function
$|\Psi\rangle$ given by eq \ref{eq-genciansatz} and the corresponding correlation energy
$\Delta E^\text{(CI)}$ as follows:
\beq
\mel{\Phi_\alpha}{H_N (1 + C^{(P_\text{A})} + C^{(P_\text{B})})}{\Phi} = 
\Delta E^\text{(CI)} \mel{\Phi_\alpha}{C^{(P_\text{A})}}{\Phi},
\label{A}
\eeq
\beq
\mel{\Phi_\beta}{H_N (1 + C^{(P_\text{A})} + C^{(P_\text{B})})}{\Phi} = 
\Delta E^\text{(CI)} \mel{\Phi_\beta}{C^{(P_\text{B})}}{\Phi},
\label{B}
\eeq
where $\ket{\Phi_\alpha} \in \mathscr{H}^{(P_\text{A})}$, $\ket{\Phi_\beta} \in \mathscr{H}^{(P_\text{B})}$,
and
\beq
\Delta E^\text{(CI)} = \mel{\Phi}{H_N (1 + C^{(P_\text{A})} + C^{(P_\text{B})})}{\Phi}
\label{C}
\eeq
(in the case of CISD calculations, where $C^{(P_\text{B})} = 0$ and the set of
determinants $\ket{\Phi_\beta}$ is empty, we only have to write eqs \ref{A} and \ref{C}). 

The main theorem of this work states that the ec-CC-I calculations, in which we obtain the $T_{3}$ and $T_{4}$
components entering eqs \ref{fcc-singles} and \ref{fcc-doubles} by defining the cluster operator $T$ as
\beq
T = \ln (1 + C^{(P_\text{A})} + C^{(P_\text{B})})
\label{t-def}
\eeq
and then solve eqs \ref{fcc-singles} and \ref{fcc-doubles} for the $T_{1}$ and $T_{2}$ clusters in the
presence of $T_{3}$ and $T_{4}$ extracted from the CI state $|\Psi\rangle$, eq \ref{eq-genciansatz},
determined by using eqs \ref{A}--\ref{C}, without eliminating any $T_{3}$ and $T_{4}$ amplitudes
resulting from the cluster analysis of $|\Psi\rangle$, return back the CI correlation energy $\Delta E^\text{(CI)}$,
eq \ref{C}, independent of truncations in $C^{(P_\text{B})}$.
In order to prove this theorem, we have to focus on the subset of CI equations
represented by eq \ref{A}, which corresponds to projections on the singly and doubly excited determinants,
\beq
\mel{\Phi_{i}^{a}}{H_N (1 + C_{1} + C_{2} + C^{(P_\text{B})})}{\Phi} =                 
\Delta E^\text{(CI)} \mel{\Phi_{i}^{a}}{C_{1}}{\Phi} ,
\label{A-singles}
\eeq
\beq
\mel{\Phi_{ij}^{ab}}{H_N (1 + C_{1} + C_{2} + C^{(P_\text{B})})}{\Phi} =
\Delta E^\text{(CI)} \mel{\Phi_{ij}^{ab}}{C_{2}}{\Phi},
\label{A-doubles}
\eeq
and
the energy formula, eq \ref{C}, which, given the absence of higher--than--two-body interactions
in the electronic Hamiltonian and the use of normal ordering in $H_{N}$, can also be written as
\beq
\Delta E^\text{(CI)} = \mel{\Phi}{H_N (C_{1} + C_{2})}{\Phi} .
\label{C-alt}
\eeq
As demonstrated, in two different ways, in Appendices A and B, the subsystem of CI equations represented
by eq \ref{A} or eqs \ref{A-singles} and \ref{A-doubles}, with the correlation energy $\Delta E^\text{(CI)}$
given by eq \ref{C} or \ref{C-alt} and the cluster operator $T$ defined by eq \ref{t-def}, so that the $T_{n}$
components with $n=1\mbox{--}4$ are obtained using eq \ref{cluster_analysis}, is equivalent to the CC amplitude
equations projected on singles and doubles, represented by eqs \ref{cceq-singles} and \ref{cceq-doubles} or,
more explicitly, \ref{fcc-singles} and \ref{fcc-doubles}.
In the proofs of this equivalence,
the CI correlation energy $\Delta E^\text{(CI)}$, eq \ref{C} or \ref{C-alt}, becomes the corresponding CC
energy given by eq \ref{fcc-energy}. The first proof of the above statement, presented in Appendix A, uses
the formal definition of $T$, eq \ref{t-def}, which brings the CI expansion, eq \ref{eq-genciansatz},
to a CC-like form given by eq \ref{eq-ccansatz}, the well-known property of the exponential ansatz that
reads (cf., e.g., refs \citenum{cizek1,cizek2,paldus-in-wilson-1992,paldus-li,leszcz,ren1,Piecuch2002})
\beq
H_N e^T \ket{\Phi} = e^T (H_N e^T)_C \ket{\Phi},
\label{CC-property}
\eeq
and the resolution of the identity in the many-electron Hilbert space,
\beq
P + P_{\rm A} + P_{\rm B} + Q = {\bf 1},
\label{resolution}
\eeq
where
\beq
P = |\Phi \rangle \langle \Phi | ,
\label{P}
\eeq
\beq
P_{\rm A} = \sum_{{\alpha}} |\Phi_{{\alpha}} \rangle
\langle \Phi_{{\alpha}} | ,
\label{PA}
\eeq
\beq
P_{\rm B} = \sum_{\beta} |\Phi_{\beta} \rangle \langle \Phi_{\beta} | ,
\label{PB}
\eeq
and
\beq
Q = \sum_{\gamma} |\Phi_{\gamma} \rangle \langle \Phi_{\gamma} |
\label{Q}
\eeq
are the projection operators on the aforementioned ${\mathscr H}^{(P)}$, $\mathscr{H}^{(P_\text{A})}$,
$\mathscr{H}^{(P_\text{B})}$, and ${\mathscr H}^{(Q)}$ subspaces and {\bf 1} is the unit operator,
to convert the CI eqs \ref{A} and \ref{C} to the CC form represented by eqs \ref{cceq-singles}, \ref{cceq-doubles},
and \ref{fcc-energy} (or \ref{fcc-singles}--\ref{fcc-energy}).
The second proof, shown in Appendix B, which relies on a diagrammatic approach, follows
the opposite direction. It starts from the CC amplitude equations projected on the singly and doubly excited
determinants, eqs \ref{fcc-singles} and \ref{fcc-doubles}, and the associated correlation energy formula,
eq \ref{fcc-energy}, which are
transformed
into the corresponding CI amplitude and energy equations,
eqs \ref{A-singles}--\ref{C-alt}, after expressing the $T_{n}$ components
with
$n=1\mbox{--}4$ in terms of the
CI
excitation operators $C_{1}$, $C_{2}$, and $C_{n}^{(P_{\rm B})}$, $n=3,4$,
using eq \ref{cluster_analysis}.

In analogy to the previously discussed case of conventional CI truncations at a given many-body rank in the
excitation operator $C$, defined by eqs \ref{eq-ciansatz} and \ref{ci-operator}, the relationship between
the ec-CC calculations based on the more general form of
the wave function $|\Psi\rangle$ defined by eq \ref{eq-genciansatz},
which encompasses a wide variety of CI approximations outside the CISD, CISDT, CISDTQ, etc. hierarchy,
and the underlying CI computations, as summarized above, has several implications.
The most apparent one is the observation that the ec-CC-I calculations based on solving
eqs \ref{fcc-singles} and \ref{fcc-doubles}, in which the $T_{3}$ and $T_{4}$ components of $T$ are obtained by cluster
analysis of the CI wave functions that describe singles and doubles fully and higher--than--double excitations
in a partial manner, return back the underlying CI energies, without any improvements,
if the purely disconnected $T_{3}$ and $T_{4}$ amplitudes
of the type of eqs \ref{t3-DC} and \ref{t4-DC}, for which
the corresponding $C_{3}^{(P_{\rm B})}$ and $C_{4}^{(P_{\rm B})}$
contributions are zero, are not eliminated. More importantly, the ec-CC-II approach, which is what one normally uses
in the CI-based ec-CC computations,
where
such purely disconnected $T_{3}$ and $T_{4}$
amplitudes are disregarded when setting up eqs \ref{fcc-singles} and \ref{fcc-doubles}, may offer improvements
over the corresponding CI calculations, but only if the triply and quadruply excited manifolds
considered in CI are incomplete, i.e., the $C_{3}^{(P_{\rm B})}$ and $C_{4}^{(P_{\rm B})}$ components
of $|\Psi\rangle$ include fractions of triples and quadruples.
Once the $C_{3}^{(P_{\rm B})}$ and $C_{4}^{(P_{\rm B})}$ operators capture all triples and quadruples
(which in practice may mean their significant fractions),
the ec-CC-I and ec-CC-II schemes based on the wave functions $|\Psi\rangle$ defined by eq \ref{eq-genciansatz}
become equivalent and the resulting ec-CC energies become identical (with the large fractions of triples
and quadruples, similar) to those obtained with CI. In other
words, assuming that singles and doubles are treated in CI fully,
the ec-CC approach can improve
the energies obtained in the underlying CI calculations only if the treatment of triples and quadruples
in the latter calculations is incomplete. In that case, after performing the ec-CC-II computations using
subsets of triples and quadruples provided by CI and selecting the $T_{3}$ and $T_{4}$ amplitudes accordingly,
to match these subsets, one can obtain the desired improvements over the corresponding CI calculations, improving
the CCSD energetics at the same time. This is especially true when the ec-CC-II energies are corrected for the
remaining $T_{3}$ (ideally, $T_{3}$ and $T_{4}$) correlations, as in the aforementioned RMRCCSD(T)
method and the CIPSI-driven ec-CC-II$_3$ approach discussed in Section \ref{section3}.
Given the fact that the likelihood of undersampling the manifolds of triples and quadruples in selected CI
calculations, represented in this work by the semi-stochastic CIPSI approach of refs
\citenum{cipsi_1,cipsi_2}, increases with the system size, the advantages of using the ec-CC-II and
ec-CC-II$_3$ methods, when compared to the underlying selected CI computations, are expected to be greater
in applications to larger problems.
On the other hand,
as already alluded to above,
the benefits offered by the CI-based ec-CC calculations compared to modern variants of selected CI techniques,
such as the semi-stochastic CIPSI method adopted in this work
or the ACI scheme of refs \citenum{adaptive_ci_1,adaptive_ci_2} used
in the recently implemented ACI-CCSD and ACI-CCSD(T) approaches, may not be as substantial as desired,
or even none, when smaller systems are examined.

Before discussing the numerical evidence supporting the key aspects of the above mathematical
analysis in Section \ref{section2.2}, it is worth mentioning that
while in this study we focus on the most popular form of the ec-CC formalism, in which one solves the CC
equations projected on singles and doubles, eqs \ref{cceq-singles} and \ref{cceq-doubles} or
\ref{fcc-singles} and \ref{fcc-doubles}, for the $T_{1}$ and $T_{2}$ clusters in the presence of
$T_{3}$ and $T_{4}$ extracted from the external, non-CC source, one can extend our considerations
to higher-order ec-CC variants that solve for higher--than--two-body components of the cluster operator $T$
(see, e.g., the ecCCSDt-CASSCF approach, discussed in ref \citenum{ec-ccsdt-cas}, for
an example).
This can be done by reusing eqs \ref{A}--\ref{C} and redefining subspaces
$\mathscr{H}^{(P_\text{A})}$ and $\mathscr{H}^{(P_\text{B})}$ and the corresponding
excitation operators $C^{(P_\text{A})}$ and
$C^{(P_\text{B})}$ that enter the CI wave function $|\Psi \rangle$
via
eq \ref{eq-genciansatz}.
To illustrate this, let us consider the CI eigenvalue problem in which
\beq
C^{(P_\text{A})} = \sum_{n=1}^{m_{A}} C_{n} ,
\label{eq-CPA-ex}
\eeq
with $m_{A} \geq 2$,
so that the corresponding subspace $\mathscr{H}^{(P_\text{A})}$ is spanned by all determinants $|\Phi_{\alpha} \rangle$
with the excitation ranks ranging from 1 to $m_{A}$, and
\beq
C^{(P_\text{B})} = \sum_{n \geq m_{A} + 1} C_{n}^{(P_\text{B})} ,
\label{eq-CPB-ex}
\eeq
where the many-body components $C_n^{(P_\text{B})}$ with $n \geq m_{A} + 1$ describe contributions from the
remaining determinants,
designated as $|\Phi_{\beta} \rangle$ and spanning
subspace $\mathscr{H}^{(P_\text{B})}$. As demonstrated in Appendix A,
the ec-CC-I calculations, in which one solves the CC system
\beq
\mel{\Phi_\alpha}{(H_N e^{T})_{C}}{\Phi} = 0 ,
\label{eq-CCA}
\eeq
where $\ket{\Phi_\alpha} \in \mathscr{H}^{(P_\text{A})}$,
for the $T_{n}$ components of $T$ with $n = 1, \ldots, m_{A}$ in the presence of the
$T_{m_{A}+1}$ and $T_{m_{A}+2}$ clusters extracted from the CI state $|\Psi\rangle$
determined by using eqs \ref{A}--\ref{C}, without eliminating any $T_{m_{A}+1}$ and $T_{m_{A}+2}$
amplitudes resulting from the cluster analysis of $|\Psi\rangle$, return back the CI
correlation energy $\Delta E^\text{(CI)}$, eq \ref{C}.
The ec-CC-II approach, where the purely disconnected $T_{m_{A}+1}$ and $T_{m_{A}+2}$
amplitudes of the type of eqs \ref{t3-DC} and \ref{t4-DC}, for which the corresponding
CI excitation coefficients in $C_{m_{A}+1}^{(P_\text{B})}$ and $C_{m_{A}+2}^{(P_\text{B})}$
are zero, are disregarded when setting up
the CC system represented by eq \ref{eq-CCA}, may improve the energies obtained in the
CI calculations used to determine $T_{m_{A}+1}$ and $T_{m_{A}+2}$, but only if the excitation
manifolds defining $C_{m_{A}+1}^{(P_\text{B})}$ and
$C_{m_{A}+2}^{(P_\text{B})}$ are not treated fully. Once all $n$-tuply excited determinants
with $n = m_{A}+1$ and $m_{A}+2$ are captured by the $C_{m_{A}+1}^{(P_\text{B})}$ and
$C_{m_{A}+2}^{(P_\text{B})}$ operators,
the ec-CC-I and ec-CC-II schemes based on the wave functions $|\Psi\rangle$ defined by eq \ref{eq-genciansatz}
become equivalent and the resulting ec-CC energies become identical to those obtained with CI.
In making the above statements, we took advantage of the fact that the $T_{n}$ clusters
with $n > m_{A} + 2$ do not enter eq \ref{eq-CCA}, since electronic Hamiltonians do
not contain higher--than--two-body interactions and the excitation ranks of determinants
$|\Phi_{\alpha} \rangle$ do not exceed $m_{A}$.

\subsection{Numerical Analysis of the ec-CC-I and ec-CC-II Schemes}
\label{section2.2}

The validity of the mathematical considerations discussed in Section \ref{section2.1} and Appendices A and B, and
of the above remarks about the anticipated accuracy patterns in the ec-CC and the underlying CI computations,
are supported by the numerical data shown in Tables \ref{table1}--\ref{table3}. Our numerical example
is the $C_{2v}$-symmetric double bond dissociation of the water molecule, as described
by the cc-pVDZ basis set,\cite{Dunning1989} in which both O--H bonds are simultaneously stretched without
changing the $\angle$(H--O--H) angle. In addition to the equilibrium geometry, designated as $R = R_{e}$,
we considered two stretches of the O--H bonds, by factors of 2 and 3, designated in our tables as
$R = 2 R_{e}$ and $3 R_{e}$, respectively. All three geometries
were
taken from ref \citenum{olsen-h2o}. Following ref \citenum{olsen-h2o}, in all of the post-SCF computations
reported in this work,
we correlated all electrons and the spherical components of $d$ functions contained in the cc-pVDZ basis
were employed throughout. In all of the CI, CC, and ec-CC computations carried out in this study, we
used the restricted Hartree--Fock (RHF) determinant as a reference $|\Phi\rangle$.

We use the $\text{H}_2 \text{O}$/cc-pVDZ system as our molecular example, since
it is small enough to allow all kinds of CI and CC computations, including high-level methods,
such as CISDTQ and beyond\cite{olsen-h2o} and CCSDTQ,\cite{olsen-h2o,Bauman2017,jed-js-pp-jcp-submitted-2021}
as well as FCI,\cite{olsen-h2o} which are all critical for the analysis of the ec-CC formalism presented in this work.
At the same time, the $C_{2v}$-symmetric double bond dissociation of the water molecule creates significant
challenges to many quantum chemistry approaches. In particular, the stretched nuclear geometries
considered in this study are characterized by substantial multi-reference correlation effects, which result in
large triply and quadruply excited CI and CC amplitudes when a single-reference framework is employed,
and which require a well-balanced description of
nondynamic and dynamic correlations (see, e.g., refs \citenum{olsen-h2o,Bauman2017}). As shown, for example,
in Table \ref{table1}, the CISDTQ approach, which is very accurate at $R = R_{e}$, recovering the
FCI energy to within a small fraction of a millihartree, struggles when the stretched geometries
are considered, increasing the errors relative to FCI to 5.819 and 16.150 millihartree at
$R = 2 R_{e}$ and $3 R_{e}$, respectively. The $R = 3 R_{e}$ geometry is so demanding that
even the CCSDTQ method, which is virtually exact at $R = R_{e}$ and $2 R_{e}$, faces a challenge, 
producing the sizable $-4.733$ millihartree error relative to FCI when the RHF reference
determinant is employed. CCSDTQ improves the erratic behavior of the CCSDT approach at $R = 3 R_{e}$,
which produces the energy 40.126 millihartree below FCI, but is not sufficient if one aims at
a highly accurate description, pointing to
the significance of higher--than--quadruply excited clusters in this case.
A similar remark applies to the CI computations, which require an explicit inclusion of
six-fold excitations if we are to recover the FCI energetics to within a millihartree at
all three geometries of water considered in this study (as can be seen in Table \ref{table1},
errors in the CISDTQP energies relative to FCI at $R = 2 R_{e}$ and $3 R_{e}$ exceed 2 and 6 millihartree,
respectively).

In performing the various ec-CC computations reported in Tables \ref{table1}--\ref{table3},
we relied on our in-house CC and cluster analysis codes, interfaced with the RHF, restricted open-shell
Hartree--Fock, and integral routines in the GAMESS package.\cite{gamess,gamess3}
The CISD, CISDT, CISDTQ, CISDTQP, and CISDTQPH wave functions, which formed the non-CC
sources of the three- and four-body clusters for the subsequent ec-CC-I and ec-CC-II calculations
based on conventional CI truncations, presented in Table \ref{table1},
were obtained using the occupation restricted multiple active space determinantal CI (ORMAS)
code\cite{Ivanic2003a,Ivanic2003b} available in GAMESS. The selected CI wave functions
used to provide the $T_{3}$ and $T_{4}$ cluster components for the ec-CC-I, ec-CC-II, and ec-CC-II$_3$
computations based on the CIPSI methodology, shown in Tables \ref{table2} and \ref{table3},
which are further elaborated on in Section \ref{section3}, were determined
with the Quantum Package 2.0 software.\cite{cipsi_1,cipsi_2}
As in the case of other post-SCF calculations reported in this study, all of our CIPSI runs
relied on the transformed one- and two-electron integrals in an RHF molecular orbital basis
generated with GAMESS. While the authors of ref
\citenum{olsen-h2o} obtained the FCI/cc-pVDZ energies of the water molecule
at $R = R_{e}$, $2 R_{e}$, and $3 R_{e}$,
we recalculated them in this study using the GAMESS determinantal FCI routines,
\cite{Ivanic2003a,Ivanic2003b,Ivanic2001}
since the FCI results at the latter two geometries
reported in ref \citenum{olsen-h2o} were not converged tightly enough. The CCSD, CCSDT,
and CCSDTQ energies were taken from ref \citenum{Bauman2017}, although we recalculated them
here as well using our in-house CC codes interfaced with GAMESS.

As shown in Table \ref{table1}, and in agreement with our mathematical analysis in Section \ref{section2.1}
and Appendices A and B,
the ec-CC-I energies obtained by solving eqs \ref{fcc-singles} and \ref{fcc-doubles} for
the singly and doubly excited clusters in the presence of the $T_{3}$ and $T_{4}$ components extracted
from the CISD, CISDT, CISDTQ, CISDTQP, and CISDTQPH wave functions, without making any {\it a posteriori}
modifications in $T_{3}$ and $T_{4}$ obtained in this way, perfectly match their CI counterparts.
This is happening, since all of the above CI truncations are characterized by a complete treatment of
the $C_{1}$ and $C_{2}$ operators. Similar is observed in the ec-CC-I computations relying on the
selected CI wave functions, obtained in this work with CIPSI, as sources of the triply and quadruply excited clusters,
when the CI diagonalization spaces are large enough to capture all or nearly all singles and doubles.
This can be seen in Tables \ref{table2} and \ref{table3}. Indeed, when the CIPSI calculations initiated
from the RHF wave functions, shown in Table \ref{table2}, capture nearly all singly and doubly excited
determinants at all three geometries of water considered in this study,
which happens when the input dimension parameter $N_\text{det(in)}$ utilized by the CIPSI
methodology to terminate the buildup of the CI diagonalization spaces, defined in Section \ref{section3},
is 100,000 or more, the resulting ec-CC-I energies match their CIPSI counterparts to within a microhartree.
When $N_\text{det(in)}$ is set at 100,000,
the final CI diagonalization spaces, which are used to obtain the wave functions that
generate the $T_{3}$ and $T_{4}$ clusters for the ec-CC computations, contain
about 200,000 $S_{z} = 0$ determinants of the $A_{1}(C_{2v})$ symmetry (see the
$N_\text{det(out)}$ values in Table \ref{table2}) and the corresponding
CIPSI runs capture about 94 \% of all singles and 98 \% of
all doubles at $R = R_{e}$, 100 \% of singles and $\sim$92 \% of doubles at $R = 2 R_{e}$, and
about 91 \% of all singly excited and 79 \% of all doubly excited determinants at $R = 3 R_{e}$.
Interestingly, the CIPSI and CIPSI-based ec-CC-I energies agree to within a millihartree
when the CI diagonalization spaces contain as little as $\sim$5,000--10,000
determinants or about 50 \% of all singles and doubles. This indicates that the CIPSI approach
is capable of correctly identifying the dominant singly and doubly excited determinants in the
early stages of the respective CI wave function buildup, so that the subsequent ec-CC-I calculations
return back the energies that are similar to those obtained in the Hamiltonian diagonalizations
used to determine $T_{3}$ and $T_{4}$.

Although
uncommon
in typical applications of
CIPSI,
we also performed a
numerical experiment in which the process of building up the CI diagonalization spaces in CIPSI runs was
initiated from the CISD wave function. We did this for the $R = 2 R_{e}$ geometry. In this case,
each CIPSI run was forced to provide a complete treatment of the $C_{1}$ and $C_{2}$ operators,
so that, based on our mathematical considerations in Section \ref{section2.1}, the resulting ec-CC-I energies and their
CIPSI counterparts should be identical. As shown in Table \ref{table3}, they are indeed identical for all
values of $N_\text{det(in)}$
that permit
CIPSI runs beginning with all singly and doubly excited determinants in the initial
diagonalization space (in the case of the all-electron calculations for
the ${\rm H_{2}O}$/cc-pVDZ system,
the CISD wave function contains 3,416 $S_{z} = 0$ determinants
of the $A_{1}(C_{2v})$ symmetry, so that $N_\text{det(in)}$ must be at least 3,416).

The above relationships between ec-CC-I and CI provide useful insights, but,
as already pointed out, the realistic applications of the CI-based ec-CC methodology
adopt the ec-CC-II protocol, where one keeps only those $T_{3}$ and $T_{4}$
amplitudes resulting from the cluster analysis of the underlying CI wave function
for which the corresponding $C_{3}$ and $C_{4}$ excitation coefficients are nonzero.
The ec-CC-II algorithm takes care of the problems resulting from the presence of the
purely disconnected forms of the $T_{3}$ and $T_{4}$ clusters, such as those
given
by eqs \ref{t3-DC} and \ref{t4-DC},
which emerge when the corresponding $C_{3}$ and $C_{4}$ amplitudes are zero,
but it does not prevent the collapse of the ec-CC energies onto their CI counterparts.
As implied by our formal analysis, the CI-based ec-CC-II calculations
can improve
the corresponding CI energetics, but in order for this to happen, the triply and quadruply excited manifolds
included in the CI diagonalizations used to determine $T_{3}$ and $T_{4}$ must be incomplete.
Otherwise, i.e., when the underlying CI computations
capture all triples and quadruples, the ec-CC-I and ec-CC-II schemes become equivalent,
returning back the 
CI
energies. 

We can see all of the above patterns in our tables.
Indeed, as demonstrated in Table \ref{table1}, variant II of the ec-CC methodology improves the CI energetics when one uses
the CISD and CISDT wave functions in the corresponding cluster analyses, but once all triples and quadruples are
included in CI, as in the case of the CISDTQ-, CISDTQP-, and CISDTQPH-based ec-CC-II calculations carried out
in this study, the ec-CC-II and the associated CI energies do not differ. This may result in unusual and
non-systematic accuracy patterns,
or even in an erratic behavior of the ec-CC-II computations. For example, normally one anticipates that when the
quality of the wave function improves the resulting energies improve as well, but this is not the case when
we examine the CI-based ec-CC-II energies of the water molecule at the stretched $R = 2 R_{e}$ geometry
shown in Table \ref{table1}. The CISD-based ec-CC-II calculation, where $C_{3} = C_{4} = 0$, so that the
$T_{3}$ and $T_{4}$ clusters entering eqs \ref{fcc-singles} and \ref{fcc-doubles} are set at
zero as well, returns the energy obtained with CCSD, reducing the massive, 72.017 millihartree, error
resulting from the CISD diagonalization to 22.034 millihartree. The ec-CC-II computation, in
which one solves eqs \ref{fcc-singles} and \ref{fcc-doubles} for $T_{1}$ and $T_{2}$ using the $T_{3}$
amplitudes extracted from the higher-rank CISDT wave function, offers further error reduction, to a 2.920
millihartree level, but the next scheme in the ec-CC-II hierarchy in Table \ref{table1}, which uses a
much better wave function in the cluster analysis than CISDT,
by returning back the CISDTQ energy, worsens the previous CISDT-based ec-CC-II result, increasing the
error by a factor of 2. The situation at $R = 3 R_{e}$ is even more peculiar. In this case,
the replacement of the CISD wave function by its higher-level CISDT counterpart in the ec-CC-II calculations
not only worsens the CISD-based ec-CC-II, i.e., CCSD energy, increasing the 10.849 millihartree unsigned error
obtained with CCSD by a factor of 7, but also places the resulting energy 77.317
millihartree below FCI. By accounting for $T_{4}$ correlations, the ec-CC-II computation employing the
CISDTQ wave function in the cluster analysis improves the erratic CISDT-based ec-CC-II result,
but since the CISDTQ-based ec-CC-II and CISDTQ energies are identical and the CISDTQ energy,
which differs from FCI by 16.150 millihartree,
is poor,
the benefits of using the ec-CC-II methodology
are none. Although the inclusion of higher--than--quadruply excited determinants, in addition to all singles
through quadruples, in the CI calculations helps (for example, the CISDTQP- and CISDTQPH-based ec-CC-II results
in Table \ref{table2}, especially the latter ones, are better than those obtained using CISDTQ), the
benefits offered by the ec-CC-II approach are none again, since the ec-CC-II and the underlying CI
computations produce identical energies once the $C_{n}$ components with $n = 1\mbox{--}4$ are treated
fully.

The above discussion
points to the need for being very careful
about evolving truncated CI wave functions used in the context of ec-CC computations. The CI algorithms
that capture the excitation spaces through quadruples, when going from one truncation level to the next
or when sampling the excitation manifolds during the CI wave function evolution,
too rapidly are not the best candidates for the ec-CC work,
since once the manifolds of singles, doubles, triples, and quadruples are saturated and all determinants
in these categories are included in the Hamiltonian diagonalization, the ec-CC calculations using $T_{3}$
and $T_{4}$ clusters extracted from such CI runs will always return back the corresponding CI energies, nothing
more. In other words, there is no point in performing the ec-CC calculations using the CI methods that saturate
the excitation spaces of singles through quadruples. This observation is independent of the quality of CI
calculations saturating these spaces. If the quality is poorer, as in the CISDTQ example discussed
above, the ec-CC computations will return back the same poorer CI energy. If the quality is good, as in the
CI calculations using all singles, doubles, triples, and quadruples and all or selected higher--than--quadruply
excited determinants, which allow the $C_{n}$ amplitudes with $n = 1\mbox{--}4$ to relax, the ec-CC
computations will still return back the energy of the underlying CI. In that case, the ec-CC energy will be
better compared to CISDTQ, where the $C_{n}$ amplitudes with $n = 1\mbox{--}4$ are not relaxed in the
presence of higher--than--quadruply excited contributions, but it will be exactly the same as that obtained
with the underlying CI, so again there is no benefit in performing the ec-CC calculations, which are adding
to the computational costs without improving the results.

In summary, based on the formal and numerical analyses presented in this section,
the truncated CI wave functions that are expected to benefit most from
the ec-CC
computations are those which attempt to probe
the many-electron Hilbert space without saturating the lower-rank excitation manifolds, especially
the excitations through quadruples, too early. We have seen this in our semi-stochastic CAD-FCIQMC work
\cite{Deustua2018,Eriksen2020}, which relies on the cluster analysis of FCIQMC wave functions, and we can see
it
in Tables \ref{table2} and \ref{table3}, where we examine the performance of the CIPSI-based ec-CC-II algorithm
and its ec-CC-II$_3$ extension that corrects the ec-CC-II energies for the missing $T_{3}$ correlations
that are not accounted for in CIPSI diagonalizations. The ec-CC-II and ec-CC-II$_3$ approaches that rely
on the CIPSI wave functions to extract the information about the leading $T_{3}$ and $T_{4}$ clusters
are discussed next.

\section{CIPSI-DRIVEN \lowercase{ec}-CC}
\label{section3}

The purpose of this section is to present and test a novel form of the ec-CC approach,
focusing on the ec-CC-II protocol and its ec-CC-II$_3$ counterpart, in which
the wave functions used to generate $T_{3}$ and $T_{4}$ clusters are obtained in the Hamiltonian
diagonalizations defining the CIPSI approach, as implemented in the Quantum Package 2.0.\cite{cipsi_1,cipsi_2}
As in the case of the numerical analysis discussed in Section \ref{section2.2},
we used the water molecule, as described by the cc-pVDZ basis set, at the equilibrium
and two displaced geometries in which both O--H bonds were simultaneously stretched by
factors of 2 and 3 without changing the $\angle$(H--O--H) angle, to illustrate the performance
of the CIPSI-driven ec-CC-II and ec-CC-II$_3$ methods.

We recall that, in analogy to many other selected CI schemes, the main idea of CIPSI is to
perform a series of CI calculations using increasingly large, iteratively defined, diagonalization spaces,
designated as $\mathcal{V}_{\text{int}}$. The construction of the $\mathcal{V}_{\text{int}}$ space for a given
CIPSI iteration is carried out using a perturbative selection of the singly and doubly excited determinants 
from the previously determined $\mathcal{V}_{\text{int}}$. To be more precise, if
$\ket{\Psi^\text{(CIPSI)}} = \sum_{\ket{\Phi_I} \in \mathcal{V}_{\text{int}}} c_I \ket{\Phi_I}$
is a CI wave function associated with a given CIPSI iteration,
where the coefficients $c_{I}$ and the corresponding energy
$E_\text{var}$ are obtained by diagonalizing
the Hamiltonian in the current $\mathcal{V}_{\text{int}}$ space,
the diagonalization space for the subsequent CIPSI iteration is constructed using a perturbative
selection of the singly and doubly excited determinants out of $\ket{\Psi^\text{(CIPSI)}}$.
Thus, if $\mathcal{V}_{\text{ext}}$ designates the space of singly and doubly excited determinants
out of $\ket{\Psi^\text{(CIPSI)}}$, we calculate the second-order MBPT energy correction associated with
each
determinant $\ket{\Phi_\rho} \in \mathcal{V}_{\text{ext}}$,
$e^{(2)}_\rho = \abs{\mel{\Phi_\rho} {H} {\Psi^\text{(CIPSI)}}}^2 / (E_\text{var} - 
\mel{\Phi_\rho} {H} {\Phi_\rho})$, and use the resulting $e^{(2)}_\rho$ values to determine how
to enlarge the current space $\mathcal{V}_{\text{int}}$.
One can initiate CIPSI iterations, each consisting of the diagonalization of the
Hamiltonian in the current space $\mathcal{V}_{\text{int}}$ to determine $\ket{\Psi^\text{(CIPSI)}}$
and the identification of the associated $\mathcal{V}_{\text{ext}}$ space needed to construct
$\mathcal{V}_{\text{int}}$ for the subsequent iteration, by starting from a single determinant,
such as the RHF wave function, or a multi-determinantal state obtained, for example, in some
preliminary CI calculation.

Following refs \citenum{cipsi_1,cipsi_2}, in the specific CIPSI model adopted in this work,
used to perform the CIPSI calculations for water reported in Tables \ref{table2} and \ref{table3},
the actual $\mathcal{V}_{\text{ext}}$
spaces were obtained by stochastic sampling of the most important singles and doubles
out of the $\ket{\Psi^\text{(CIPSI)}}$ wave functions and
the sampled determinants $\ket{\Phi_\rho}$ generated in each CIPSI iteration were arranged
in descending order according to their $|e^{(2)}_\rho|$ values. We then enlarged
each space $\mathcal{V}_{\text{int}}$, to be used in the subsequent CI diagonalization,
starting from the determinants $\ket{\Phi_\rho}$ with the largest $|e^{(2)}_\rho|$
contributions and moving toward those with the smaller values of $|e^{(2)}_\rho|$,
until the dimension of $\mathcal{V}_{\text{int}}$ was increased by the user-defined factor $f$ (in reality,
this increase in the dimension of $\mathcal{V}_{\text{int}}$, from one CIPSI iteration to the next,
was always slightly larger to ensure that the resulting CI wave function
remained an eigenfunction of total spin). In all of the CIPSI calculations
presented in Tables \ref{table2} and \ref{table3},
we set $f$ at
2, which
is the default value of $f$ in Quantum Package
2.0, forcing
the CIPSI wave function $\ket{\Psi^\text{(CIPSI)}}$ to grow in a tempered manner,
without saturating the lower-rank excitation manifolds too rapidly, while probing
the many-electron Hilbert space more effectively at the same time.
We experimented with various choices of $f$ in the calculations for the ${\rm H_{2}O}$/cc-pVDZ system examined
in this work, and we will return to the investigation of the relationship between the value of $f$ and the
accuracy of the CIPSI-based ec-CC-II and ec-CC-II$_{3}$ energies using a larger number of molecules and
basis sets in the future study, but our preliminary analysis in the Supporting Information
clearly shows that neither the rapid growth of the CIPSI wave functions by using $f \gg 2$
nor the ultra-slow wave function growth corresponding to $f$ only slightly above 1
benefit the ec-CC-II and ec-CC-II$_{3}$ computations using CIPSI. One might argue that using the values
of $f$ slightly larger than 2 can help by reducing the number of Hamiltonian diagonalizations
in the CIPSI runs, while maintaining the accuracy of the ec-CC-II and ec-CC-II$_{3}$ energies obtained with
$f = 2$, but based on the data shown in Table S1 of the
Supporting Information, the default dimension-doubling mechanism adopted in the implementation of CIPSI
in Quantum Package 2.0 and in the CIPSI-driven ec-CC-II and ec-CC-II$_{3}$ calculations reported in
Tables \ref{table2} and \ref{table3} is certainly a good choice.

To obtain the final
CI wave function of a given CIPSI run used to determine the $T_{3}$ and $T_{4}$ clusters
employed in the ec-CC computations, we chose to terminate the
sequence of CIPSI diagonalizations when the dimension of space $\mathcal{V}_{\text{int}}$
exceeded the input parameter $N_\text{det(in)}$. Due to the
aforementioned dimension-doubling growth mechanism, the size of the CI wave function at the
end of a given CIPSI calculation, denoted as $N_\text{det(out)}$, always exceeded
$N_\text{det(in)}$, but never by more than a factor of 2. 
As already alluded to above, the CI wave functions generated in our CIPSI runs,
including the final ones used in the cluster analysis prior to the ec-CC work,
were eigenfunctions of the total
spin $S^{2}$ and $S_{z}$ operators. In the RHF-based computations
for water reported in this article, they were all singlet states.

As a byproduct of calculating
energy corrections $e^{(2)}_\rho$ associated with the sampled determinants
$\ket{\Phi_\rho} \in \mathcal{V}_\text{ext}$ generated in each CIPSI iteration,
in addition to the variational energies $E_\text{var}$, one has an immediate
access to the total second-order multi-reference MBPT corrections $\Delta E^{(2)} = 
\sum_{\ket{\Phi_\rho} \in \mathcal{V}_\text{ext}} e_{\rho}^{(2)}$.
Thus, for
each CIPSI run carried out in this
study, we
report the uncorrected energy $E_\text{var}$ corresponding to the CI wave function obtained
in the last Hamiltonian diagonalization of that run and its perturbatively corrected
$E_\text{var} + \Delta E^{(2)}$ counterpart.
We also report the $E_\text{var} + \Delta E_{r}^{(2)}$
energies, where $\Delta E_{r}^{(2)}$ represents the renormalized form of the second-order
MBPT correction introduced in ref \citenum{cipsi_2}. With an exception of the very small
CIPSI diagonalization spaces, in the calculations reported in this work,
the two perturbative corrections to the variational energies $E_\text{var}$
produce nearly identical results, so in the discussion below we focus on the
$E_\text{var} + \Delta E^{(2)}$ energies.
The determination of $\Delta E^{(2)}$ is also important for a different reason.
Although the stopping criterion adopted in our CIPSI runs executed with Quantum Package 2.0
relies on the above wave function termination parameter $N_\text{det(in)}$, the CIPSI iterations can
also stop if the magnitude of the total second-order MBPT correction $\Delta E^{(2)}$ falls below a threshold
parameter $\eta$. To prevent this from happening, we used a very tight $\eta$ value of $10^{-6}$ hartree.

After the completion of each CIPSI run, we cluster analyzed the wave function $\ket{\Psi^\text{(CIPSI)}}$
obtained in the final Hamiltonian diagonalization of that run and used the resulting $T_{3}$ and $T_{4}$
components to perform the corresponding ec-CC computations. Since in the ec-CC-II calculations
discussed
in this section the purely disconnected $T_{3}$ and $T_{4}$ amplitudes of the type of eqs
\ref{t3-DC} and \ref{t4-DC}, for which the corresponding $C_{3}$
and $C_{4}$ contributions are zero, are disregarded when setting up eqs \ref{fcc-singles} and \ref{fcc-doubles},
we corrected the ec-CC-II correlation energies $\Delta E$, determined using eq \ref{fcc-energy},
for the missing $T_{3}$ effects
not captured by ec-CC-II. In order to do this, we adopted the formulas that we previously used to develop the
deterministic\cite{jspp-chemphys2012,jspp-jcp2012,Bauman2017} and semi-stochastic
\cite{stochastic-ccpq-prl-2017,stochastic-ccpq-molphys-2020,jed-js-pp-jcp-submitted-2021}
CC($P$;$Q$) approaches. According to the biorthogonal moment expansions behind the CC($P$;$Q$)
framework, the corrections to the ec-CC-II correlation energies $\Delta E$ due to the $T_{3}$
effects not captured by the CIPSI wave functions,
subjected to the so-called two-body approximation introduced in ref \citenum{jspp-chemphys2012}
that has been shown to provide a highly accurate
representation of these effects,\cite{jspp-chemphys2012,jspp-jcp2012,Bauman2017}
can be defined as follows:
\beq
\delta_{3} = \sum_{
|\Phi_{ijk}^{abc}\rangle \not\in \mathcal{V}_{\text{int}}}
{{\ell}}_{ijk}^{abc}(2) \: {\mathfrak M}^{ijk}_{abc}(2) .
\label{delta-triples}
\eeq
Here,
\beq
{\mathfrak M}_{abc}^{ijk}(2) =
\langle \Phi_{ijk}^{abc} | \overline{H_{N}}(2) |\Phi \rangle
\label{eq-mom-explicit}
\eeq
are the moments of CC equations corresponding to projections on the triply excited determinants
missing in the final CI diagonalization space $\mathcal{V}_{\text{int}}$ of a given CIPSI run,
where the similarity-transformed Hamiltonian
\beq
\overline{H_{N}}(2) = (H_{N} e^{T_{1}+T_{2}})_{C} = e^{-T_{1}-T_{2}} H_{N} e^{T_{1}+T_{2}}
\label{eq-hbar-mod}
\eeq
is obtained using the singly and doubly excited clusters resulting from the ec-CC-II calculations.
The $\ell_{ijk}^{abc}(2)$ coefficients, which are given by the expression
\beq
\ell_{ijk}^{abc}(2)
= \langle \Phi | (\Lambda_{1} + \Lambda_{2}) \: \overline{H_{N}}(2) |\Phi_{ijk}^{abc}\rangle/
(\Delta E - \langle \Phi_{ijk}^{abc} | \overline{H_{N}}(2) | \Phi_{ijk}^{abc}\rangle) ,
\label{eq-ell}
\eeq
are the deexcitation amplitudes that require solving the companion left CC equations
for the one- and two-body components of the operator $\Lambda$ that generates the CC bra state
$\langle \Phi | (1 + \Lambda) e^{-T}$ (cf., e.g., ref \citenum{Bartlett2007a}), i.e.,
\beq
\langle \Phi|(1 + \Lambda_{1} + \Lambda_{2}) [\overline{H_{N}}(2) - \Delta E \, {\bf 1}]|\Phi_{\alpha}\rangle = 0 ,
\label{eq-leftcc}
\eeq
where $|\Phi_{\alpha}\rangle$ represents the singly and doubly excited determinants.

As already alluded to in Section \ref{section2}, the ec-CC approach, in which the
ec-CC-II correlation energy $\Delta E$, eq \ref{fcc-energy}, is corrected for
the missing $T_{3}$ effects not captured by the underlying CI calculations using
correction $\delta_{3}$ given by eq \ref{delta-triples}, defines the ec-CC-II$_3$ method.
We use the above correction $\delta_{3}$ rather than its simplified CCSD(T)-like analog
adopted in the RMRCCSD(T) and ACI-CCSD(T) work,\cite{LiPaldus2006,Aroeira2020} since it is well established
that, in analogy to the completely renormalized CC approaches, such as
CR-CC(2,3),\cite{crccl_jcp,crccl_cpl,crccl_molphys,crccl_jpc}
the CC($P$;$Q$) moment corrections are a lot more robust.
\cite{jspp-chemphys2012,jspp-jcp2012,Bauman2017} Furthermore, there are molecular applications
where the CCSD(T)-type corrections
to the CC calculations including subsets of connected triple excitations do not properly
account for the missing $T_{3}$ correlations; instead
of improving the underlying
CC results, they
make the results worse (see, e.g., ref \citenum{jspp-jcp2012} for a discussion).
It should also be 
mentioned
that the computational cost of determining the $\delta_{3}$
correction to the ec-CC-II correlation energy is only twice the cost of calculating
the corresponding (T) correction, i.e., the replacement of $\delta_{3}$, eq \ref{delta-triples},
by its (T) counterpart 
would offer
no significant computational benefits either.
One might contemplate using other forms of the triples corrections that utilize the $\Lambda$ operator
of the CC theory, such as those proposed in refs
\citenum{stanton1997,crawford1998,ref:26,gwaltney1,gwaltney3,eomccpt,ccsdpt2,bartlett2008a,bartlett2008b,%
eriksen1,eriksen2},
or the older moment corrections preceding CR-CC(2,3) and its CC($P$;$Q$) generalization,
developed in refs \citenum{leszcz,ren1} and adopted
in the recent DMRG- and HCI-based ec-CC work,\cite{chan2021ecCC} but, judging by the superior performance
of the CR-CC(2,3) approach in comparison with other triples corrections (see, e.g., refs
\citenum{crccl_jcp,crccl_cpl,crccl_molphys,crccl_jpc,taube2010,Magoulas2018,Yuwono2019})
and further improvements over CR-CC(2,3) offered by the CC($P$;$Q$) framework,
especially in multi-reference situations where the coupling of the lower-rank ($T_{1}$ and $T_{2}$)
and higher-rank ($T_{3}$ and $T_{4}$) clusters becomes significant,
\cite{jspp-chemphys2012,jspp-jcp2012,Bauman2017,stochastic-ccpq-prl-2017,jed-js-pp-jcp-submitted-2021}
the $\delta_{3}$ correction defined by eq \ref{delta-triples} represents one of the best choices
one can make to correct the ec-CC-II energy for the missing $T_{3}$ correlations not captured by the
underlying CIPSI computations. While it may be interesting to compare the ec-CC-II$_3$ results obtained
using different forms of the triples corrections to the ec-CC-II energies, and we plan to return to this
topic in the future study, in the initial implementation of the CIPSI-driven ec-CC-II$_3$ methodology
explored in this work we focus on the $\delta_{3}$ correction given by eq \ref{delta-triples}.

At this time, our deterministic ec-CC-II and ec-CC-II$_3$ codes, along with the routines that allow one to
perform the corresponding ec-CC-I calculations examined in Section \ref{section2.2}, are capable of reading
the non-CC wave functions for the subsequent cluster analysis and the CC computations based on eqs
\ref{fcc-singles}--\ref{fcc-energy} and \ref{delta-triples}--\ref{eq-leftcc} from GAMESS (the
determinantal ORMAS\cite{Ivanic2003a,Ivanic2003b} and FCI\cite{Ivanic2003a,Ivanic2003b,Ivanic2001}
runs) and Quantum Package 2.0
(CIPSI calculations\cite{cipsi_1,cipsi_2}). We also have the semi-stochastic variant of the ec-CC codes
that can work with the wave functions obtained with FCIQMC, as in the CAD-FCIQMC methodology
introduced in ref \citenum{Deustua2018} and further elaborated on in the Supporting Information to
ref \citenum{Eriksen2020}.
The remaining information about our ec-CC codes used in this work can be found in Section \ref{section2.2}.

We now move to the discussion of our CIPSI-based all-electron ec-CC-II and ec-CC-II$_3$ calculations
for the $C_{2v}$-symmetric double bond dissociation of the water molecule, as described by the
cc-pVDZ basis set, summarized in Tables \ref{table2} and \ref{table3}. The information about the
nuclear geometries used in these calculations has been provided in Section \ref{section2.2}. In analogy
to the ec-CC-I computations discussed in Section \ref{section2.2}, we performed two sets
of CIPSI-driven ec-CC-II and ec-CC-II$_3$ calculations. In the first set, summarized in
Table \ref{table2}, where we applied the CIPSI-based ec-CC-II and ec-CC-II$_3$ approaches to
the equilibrium geometry, $R = R_{e}$, and two stretches of both O--H bonds, including
$R = 2 R_{e}$ and $3 R_{e}$, we initiated each CIPSI run from the corresponding RHF wave functions.
In the second set, summarized in Table \ref{table3}, where we focused on
the $R = 2 R_{e}$ geometry,
we forced each CIPSI calculation preceding the
ec-CC-II and ec-CC-II$_3$ steps to provide a complete treatment of the $C_{1}$ and $C_{2}$ operators
by initiating the CIPSI runs from the wave function obtained with CISD. In each case,
we considered multiple values of the input parameter $N_\text{det(in)}$ used to terminate
the CIPSI runs by sampling $N_\text{det(in)}$ in a semi-logarithmic manner.
In the case of the CIPSI calculations initiated from the single-determinantal RHF
wave function (Table \ref{table2}), the smallest value of $N_\text{det(in)}$ was 1.
As already mentioned in Section \ref{section2.2}, the smallest value of $N_\text{det(in)}$
used in the CIPSI calculations initiated from the CISD state had to be at least 3,416
(the number of $S_{z} = 0$ determinants defining the CISD ground-state problem if
the  $A_{1}(C_{2v})$ symmetry is employed). The smallest value of $N_\text{det(in)}$
that we used in this case was 5,000.

The results in Tables \ref{table2} and \ref{table3} demonstrate that both ec-CC-II and ec-CC-II$_3$,
especially the latter method, offer significant improvements over the underlying CIPSI computations.
This is particularly true when the relatively small values of $N_\text{det(in)}$ and the similarly small
CIPSI diagonalization spaces, as defined by $N_\text{det(out)}$, are employed. They also significantly
improve the corresponding CCSD results, where $T_{3}$ and $T_{4}$ are assumed to be zero. As shown in
Table \ref{table2},
with as little as about 5,000--6,000 determinants in the CIPSI calculations initiated from RHF, which
capture approximately 40--50 \% of singles, 20--70 \% of doubles, 1--2 \% of triples, and 0.2 \% of
quadruples, the ec-CC-II approach reduces the approximately 8, 14, and 10 millihartree errors relative
to FCI obtained with CIPSI at $R = R_{e}$, $2R_{e}$, and $3R_{e}$, respectively, to
about 2--4 millihartree. The triples correction $\delta_{3}$, eq \ref{delta-triples},
used in the ec-CC-II$_{3}$ calculations,
reduces these errors even further, to 0.012, 0.226, and 1.507 millihartree, respectively.
These are impressive improvements, especially if we realize that the CISD and CCSD
calculations, which use the same numbers of singly and doubly excited amplitudes as
ec-CC-II and only slightly smaller numbers of excitation amplitudes than the $N_\text{det(in)} = 5,000$
CIPSI runs, produce much larger errors, which are more than 12 millihartree at $R = R_{e}$,
more than 72 millihartree at $R = 2R_{e}$, and almost 165 millihartree at $R = 3R_{e}$
in the case of CISD and about 4, 22, and 11 millihartree, respectively, when the CCSD approach
is employed (cf. Table \ref{table1}).
As a matter of fact, by comparing the results in Tables \ref{table1} and \ref{table2},
we can see that the $N_\text{det(in)} = 5,000$ ec-CC-II$_{3}$ calculations
initiated from RHF are considerably more accurate
than the CCSDT, CISDTQ, and even CISDTQP calculations, which are more expensive by
orders of magnitude and which use 90,279, 1,291,577,
and 10,502,233 $S_{z} = 0$ excitation amplitudes, when the $A_{1}(C_{2v})$ symmetry is employed,
as opposed to 3,145 amplitudes used by the underlying ec-CC-II approach and about 5,000--6,000 determinants
included in the CIPSI diagonalizations needed to extract the $T_{3}$ and $T_{4}$
clusters for the ec-CC computations. Interestingly, while the CIPSI-driven ec-CC-II$_{3}$ method
using $N_\text{det(in)} = 5,000$ is less accurate than the CCSDTQ approach in the
$R = R_{e}\mbox{--}2R_{e}$ region, it becomes competitive with it when the largest stretch of both O--H bonds
in water considered in this study, i.e., $R = 3R_{e}$, is examined, reducing the
$-4.733$ millihartree error relative to FCI obtained with CCSDTQ by more than a factor of 3.
Compared to ec-CC-II,
there is an extra cost associated with the determination of the triples correction
$\delta_{3}$ in the ec-CC-II$_{3}$ calculations, but the computational costs associated
with this noniterative correction, being similar
to those of CCSD(T), are less than the cost of a single iteration of CISDT or CCSDT. One also
has to keep in mind that, in analogy to CCSD(T), one does not have to store higher--than--two-body quantities
when forming the triples correction defined by eqs \ref{delta-triples}--\ref{eq-leftcc}.

The results of the ec-CC-II and ec-CC-II$_{3}$ calculations employing the $T_{3}$ and $T_{4}$
amplitudes extracted from the CIPSI wave functions become even more accurate when the
CIPSI diagonalization spaces start growing. For example, when the wave function termination
parameter $N_\text{det(in)}$ is set at 50,000, which translates into about 80,000--90,000 determinants
participating in the final Hamiltonian diagonalizations of the corresponding CIPSI runs, 
the ec-CC-II calculations reduce the 2.612, 2.436, and 0.906 millihartree errors relative to FCI
at $R = R_{e}$, $2R_{e}$, and $3R_{e}$, respectively, obtained with CIPSI,
by factors of 3--4, to 0.626 millihartree at $R = R_{e}$,
0.788 millihartree at $R = 2R_{e}$, and 0.341 millihartree at $R = 3R_{e}$. The $\delta_{3}$
correction reduces the already small errors obtained with the CIPSI-driven ec-CC-II
approach at $R = R_{e}$ and $2R_{e}$ even further, to 0.168 and 0.515 millihartree, respectively.
The ec-CC-II$_{3}$ calculations do not improve the underlying ec-CC-II result at $R = 3R_{e}$ any longer,
possibly because the $\delta_{3}$ correction defined by eq \ref{delta-triples}
takes care of only the missing $T_{3}$ correlations not captured
by CIPSI, without correcting the ec-CC-II energies for the missing $T_{4}$ effects,
which at $R = 3R_{e}$ may become substantial (cf., e.g., the large difference between the CCSDTQ and
CCSDT energies in Table \ref{table1}). On the other hand,
the 0.358 millihartree error obtained with ec-CC-II$_{3}$ in a highly challenging multi-reference
situation created by the $R = 3R_{e}$ structure of the water molecule is
a very accurate result.
One has to keep in mind that the much more expensive CISDTQ and CCSDTQ methods, which use 1,291,577
excitation amplitudes, as opposed to $N_\text{det(out)} = 92,707$ determinants
in the last CIPSI diagonalization space corresponding
to $N_\text{det(in)} = 50,000$ and 3,145 singles and doubles participating in the ec-CC steps,
combined with the relatively inexpensive noniterative correction $\delta_{3}$,
produce the 16.150 and $-4.733$ millihartree errors, respectively,
when the $R = 3R_{e}$ geometry is considered.
Similar to the
$N_\text{det(in)} = 5,000$ case, the $N_\text{det(in)} = 50,000$ CIPSI-based
ec-CC-II$_{3}$ calculations are also more accurate than CISDTQP, which uses
10,502,233 excitation amplitudes.

When we look at the overall picture emerging from the results reported in Tables \ref{table2} and \ref{table3},
it is quite clear that the CIPSI-driven ec-CC-II and ec-CC-II$_{3}$ computations, especially the latter ones,
offer a rapid convergence toward FCI with the relatively small CIPSI diagonalization spaces.
As shown, for example, in Tables \ref{table2} and \ref{table3}, when one uses
about 1,000,000 determinants in the final diagonalizations of the CIPSI runs, both the
uncorrected ec-CC-II and the corrected ec-CC-II$_{3}$ calculations recover the FCI energetics at all
three geometries of the water molecule examined in this work, including the most challenging
$R = 3R_{e}$ structure, to within 0.1 millihartree. To appreciate this result, one has to keep in mind that 1,000,000
determinants in the diagonalization space is nowhere near the dimension of the FCI ground-state problem, which is
451,681,246 if the $S_{z} = 0$ $A_{1}(C_{2v})$-symmetric determinants are considered.
What certainly helps the ec-CC-II and ec-CC-II$_{3}$ calculations in achieving this remarkable performance
is the aforementioned tempered growth of the wave function in the consecutive CIPSI iterations, which allows the CIPSI
algorithm
to efficiently sample the
many-electron Hilbert space, without saturating the lower-rank excitation manifolds, especially
the excitations through quadruples, too early,
while relaxing the singly through quadruply excited CI amplitudes, used to determine $T_{3}$ and $T_{4}$
clusters, in the presence of higher-order correlations.
As already demonstrated in Section \ref{section2},
mathematically and numerically, if CIPSI saturated the lower-rank excitation manifolds too rapidly,
without bringing information about higher--than--quadruply excited contributions, our ec-CC-II
computations would collapse onto the results of the respective Hamiltonian diagonalizations. This emphasizes the
significance of the appropriate design of the CI (in general, non-CC) methodologies used to
provide information about the $T_{3}$ and $T_{4}$ clusters in ec-CC considerations. Our
computations suggest that the current design of the CIPSI algorithm in Quantum Package 2.0
is well suited for the ec-CC-II and ec-CC-II$_{3}$ approaches developed in this study.

Having stated all of the above, one cannot ignore the fact
that the CIPSI approach can be very efficient in its
own right, especially when the variational energies $E_\text{var}$ resulting from the underlying
CI diagonalizations are corrected for the
leading remaining correlations using the
second-order multi-reference MBPT
corrections $\Delta E^{(2)}$. As shown, for example, in 
Table \ref{table2}, when the final diagonalizations of the CIPSI runs involve about
1,000,000 determinants, the $E_\text{var} + \Delta E^{(2)}$ energies are within a few microhartree
from FCI, independent of the nuclear geometry, despite the fact that the FCI space is about 500 times larger.
When $N_\text{det(in)} = 1,000,000$, which translates into diagonalization spaces on the order
of 1.2--1.4 million, the $E_\text{var} + \Delta E^{(2)}$ energies reported in Table \ref{table2} are
superior to their already very accurate
ec-CC-II and ec-CC-II$_{3}$ counterparts.
Even with as little as about 80,000--90,000 determinants
in the final diagonalizations of the CIPSI runs, which result from setting the
$N_\text{det(in)}$ parameter at 50,000, the $E_\text{var} + \Delta E^{(2)}$ energies
are still within 0.1 millihartree from FCI.
While the perturbatively corrected CIPSI calculations for larger many-electron systems
and larger basis sets may require additional extrapolations to achieve similarly
accurate results,\cite{cipsi_2,cipsi_benzene} the fact of the matter remains that CIPSI
represents a powerful computational tool capable of generating high-quality
results by itself. The CIPSI-driven ec-CC-II and ec-CC-II$_{3}$ approaches
are capable of substantially improving the purely variational
CIPSI energies and the results of lower-rank CC calculations,
but one has to keep in mind that CIPSI and other modern variants of selected CI techniques
can be made very accurate
too.

\section{CONCLUSIONS}
\label{section4}

One of the most interesting ways of extending the applicability of single-reference CC approaches
to multi-reference and strongly correlated systems is offered by the ec-CC formalism. The key idea
of all ec-CC methods is to solve the CC equations for the lower-rank cluster components, such as
$T_{1}$ and $T_{2}$, in the presence of their higher-order $T_{n}$ counterparts
(typically, $T_{3}$ and $T_{4}$) extracted from a non-CC source that behaves well in situations
characterized by stronger nondynamic correlations. In this paper, we have focused on the ec-CC
methods, in which one solves the CC equations projected on the singly and doubly excited
determinants, eqs \ref{fcc-singles} and \ref{fcc-doubles}, respectively,
for the $T_{1}$ and $T_{2}$ clusters using the $T_{3}$ and $T_{4}$
contributions obtained via cluster analysis of truncated CI wave functions.

The present study has had two main objectives. The first objective has been a thorough examination
of the mathematical content of the ec-CC equations, backed by the appropriate numerical
analysis, in which we have attempted to identify the truncated CI states that,
after extracting the $T_{n}$ components of the cluster operator $T$ with $n = \mbox{1--4}$ from
them via the cluster analysis procedure adopted in all ec-CC considerations, satisfy eqs
\ref{fcc-singles} and \ref{fcc-doubles}. This is an important topic, since, by solving
eqs \ref{fcc-singles} and \ref{fcc-doubles} for the $T_{1}$ and $T_{2}$ clusters
in the presence of the $T_{3}$ and $T_{4}$ amplitudes extracted from such states, the
ec-CC procedure can only return back the corresponding CI energies, without improving them at all.
The second objective has been the
initial
exploration of a novel flavor of the ec-CC approach in which
$T_{3}$ and $T_{4}$ contributions are
obtained by the cluster analysis of the truncated CI wave functions resulting from one of the
most successful selected CI methods abbreviated as CIPSI.

We have demonstrated that the ec-CC calculations performed by solving eqs \ref{fcc-singles} and \ref{fcc-doubles},
where the $T_{3}$ and $T_{4}$ components are obtained by cluster analysis of the CI
wave functions that describe singles and doubles fully and higher--than--double excitations
in a complete or partial manner and where all $T_{3}$ and $T_{4}$ amplitudes generated in this
way are kept, as in the ec-CC-I protocol discussed in Section \ref{section2}, return back
the underlying CI energies. This means that the ec-CC computations, which use the wave
functions obtained with the conventional CI truncations, such as CISD, CISDT, CISDTQ, etc.,
or with any other CI method that provides a complete treatment of the single and double
excitation manifolds, offer no improvements over CI if no {\it a posteriori} modifications
are made in $T_{3}$ and $T_{4}$ extracted from CI. In reality, i.e., in typical
applications of the CI-based ec-CC methodology, one disregards the purely disconnected
$T_{3}$ and $T_{4}$ amplitudes of the type of eqs \ref{t3-DC} and \ref{t4-DC}, for which
the corresponding CI excitation coefficients are zero, as in the ec-CC-II algorithm
discussed in Sections \ref{section2} and \ref{section3}, but this does not prevent the
collapse of the resulting ec-CC energies onto their CI counterparts. As shown in this
work,
the
ec-CC-II approach may offer improvements
over the underlying CI calculations, but only if the triply and quadruply excited manifolds
considered in CI are incomplete. Once the CI calculation captures all triples and quadruples,
as in CISDTQ, CISDTQP, CISDTQPH, etc., or in any other CI truncation that treats singles through
quadruples fully, the ec-CC-I and ec-CC-II
schemes
become
equivalent and the ec-CC computations offer no benefits compared to CI, while adding to the computational cost.
In other words, in order for the ec-CC computations based on eqs \ref{fcc-singles} and \ref{fcc-doubles}
to significantly improve the energetics obtained with the underlying CI approach, it is essential to avoid
a complete or nearly complete treatment of the triple and quadruple excitation manifolds.
In that case, after solving eqs \ref{fcc-singles} and \ref{fcc-doubles} for $T_{1}$ and $T_{2}$
in the presence of $T_{3}$ and $T_{4}$ extracted from CI, in which
the amplitudes that do not have the companion triple and quadruple excitation CI coefficients
are ignored, it is useful to correct the resulting ec-CC-II energies for the
remaining $T_{3}$ and $T_{4}$ or at least $T_{3}$ correlations, as in the RMRCCSD(T)
approach of ref \citenum{LiPaldus2006} or the CIPSI-driven ec-CC-II$_3$ method
introduced in this work, to name representative examples. We have also considered
higher-order ec-CC-I and ec-CC-II variants that solve for higher--than--two-body
components of the cluster operator $T$ and examined their relations with the underlying CI approaches.

Our mathematical and numerical analyses imply that the truncated CI wave functions that
are best suited for the ec-CC computations are those
that
efficiently
sample the many-electron Hilbert space, without saturating the lower-rank excitation manifolds, especially
the excitations through quadruples, too rapidly,
while adjusting the singly through quadruply excited CI amplitudes to the dominant higher--than--quadruply
excited contributions.
As shown in this work,
the modern formulation of the CIPSI approach, developed in refs \citenum{cipsi_1,cipsi_2},
which achieves a tempered growth of the wave function through
a systematic sequence of CI calculations, combined with perturbative and stochastic
analyses of the excitation spaces used in Hamiltonian diagonalizations, is capable
of providing such CI states. By examining the $C_{2v}$-symmetric double bond dissociation
of the water molecule, including the very challenging region where both O--H bonds
are stretched by a factor of 3, so that even the sophisticated levels of the CC theory,
such as full CCSDT and CCSDTQ, struggle, we have demonstrated that the CIPSI-based
ec-CC-II method, described in Section \ref{section3}, is capable of providing highly accurate
results with the relatively low computational costs,
especially when the ec-CC-II energies are corrected for the missing $T_{3}$ correlations
via the ec-CC-II$_3$ scheme. The CIPSI-driven ec-CC-II$_3$ energies are so accurate that they
are competitive with those obtained with the much more expensive
high-level CC and CI methods, such as CCSDTQ, CISDTQP, or even CISDTQPH.
Most remarkably, the ec-CC-II and ec-CC-II$_{3}$ computations, especially the latter ones,
offer a rapid convergence toward FCI, reaching submillihartree accuracies
with the relatively small CIPSI diagonalization spaces used to determine $T_{3}$ and $T_{4}$.
The fast convergence of the energies obtained in the CIPSI-enabled ec-CC-II runs toward FCI is reminiscent of
the FCIQMC-enabled ec-CC computations using the CAD-FCIQMC method, observed in refs
\citenum{Deustua2018,Eriksen2020}. This only reinforces our view that the CI approaches
capable of efficiently sampling the many-electron Hilbert space through a tempered
evolution of the wave function, without populating the lower-rank excitation manifolds, especially
the excitations through quadruples, too fast, benefit the ec-CC computations most.

While our initial tests of the CIPSI-driven ec-CC-II and ec-CC-II$_{3}$ approaches reported
in this study are very promising, encouraging us to continue our work in this direction,
the present study has also pointed out that the underlying CIPSI method,
especially when one adds the second-order multi-reference MBPT
corrections to the variational energies obtained in the CIPSI Hamiltonian diagonalizations,
can be made very accurate as well, being competitive with
or even more accurate than
the ec-CC-II$_{3}$ results.
This is not a criticism of the idea of ec-CC, but, rather, a recognition of the fact that
the new generations of selected CI techniques, such as those developed in refs
\citenum{adaptive_ci_1,adaptive_ci_2,asci_1,asci_2,ici_1,ici_2,shci_1,shci_2,shci_3,cipsi_1,cipsi_2},
the stochastic CI approaches,
especially FCIQMC,\cite{Booth2009,Cleland2010,Ghanem2019,ghanem_alavi_fciqmc_2020,ghanem_alavi_fciqmc_2021}
and other methods that utilize CI concepts, such as the correlation energy extrapolation by intrinsic scaling
\cite{ceeis1,ceeis2,ceeis3} and the
incremental\cite{zimmerman1,zimmerman2}
and many-body expanded\cite{eriksengauss1,eriksengauss2,eriksengauss3} FCI,
along with non-traditional alternatives to FCI, such as DMRG,
\cite{dmrg1,dmrg2,dmrg3,dmrg4,dmrg5,dmrg6,dmrg7}
have become highly competitive with the best CC solutions (cf., e.g.,
refs \citenum{Eriksen2020,cipsi_benzene,cipsi_excbench_1,cipsi_excbench_2,cipsi_excbench_3,%
cipsi_excbench_5,shci_excbench,shci_cr2bench,shci_gaussian2bench}
for selected recent examples; see, also, refs \citenum{cipsi_excbench_4,eriksen-review-2021,dmrg8}
for recent perspectives).

It would be important to investigate if our initial observations regarding
the performance of the CIPSI-based ec-CC-II and ec-CC-II$_{3}$ approaches reported in this study remain
true in a wider range of molecular applications. Furthermore, it would be interesting to examine if
the ec-CC approaches based on other selected CI
algorithms developed in recent
years, especially the semi-stochastic CI approaches enabling
efficient
sampling of the many-electron Hilbert space, such as the
HCI
framework of refs \citenum{shci_1,shci_2,shci_3}
used in the ec-CC study reported in ref \citenum{chan2021ecCC},
are as accurate and as efficient as
our
CIPSI-enabled ec-CC-II and ec-CC-II$_{3}$
schemes.
It would also be important to investigate if our ec-CC-II$_{3}$ results could further be
improved by correcting the CIPSI-driven ec-CC-II energies for the missing $T_{3}$ as well as $T_{4}$
correlation effects not captured during CIPSI runs, rather than the missing $T_{3}$ correlations only.
In analogy to the triples correction $\delta_{3}$ adopted in this work, we could take
advantage of the formulas that we previously used to develop and benchmark the CC($P$;$Q$)
approaches correcting the active-space CC energies, such as CCSDtq, for the missing triples and quadruples.
\cite{jspp-chemphys2012,Bauman2017,Magoulas2018,Yuwono2019}
Such corrections might also benefit the aforementioned semi-stochastic CAD-FCIQMC methodology
based in the cluster analysis of FCIQMC wave functions, which we
have been pursuing in parallel with the ec-CC development work reported in this article.

\section*{Supporting Information}

The results of the all-electron CIPSI calculations, initiated from the RHF wave function, and of the corresponding
ec-CC computations for the stretched, $R = 2 R_{e}$, structure of ${\rm H_{2}O}$,
as described by the cc-pVDZ basis set, in which the wave function termination input parameter $N_\text{det(in)}$
employed by the CIPSI algorithm was fixed at 100,000 and the parameter $f$ that controls the growth of the
Hamiltonian diagonalization spaces was varied from 1.05 to 10.
This information is available free of charge via the Internet at http://pubs.acs.org.

\begin{acknowledgement}
This work has been supported by the Chemical Sciences, Geosciences
and Biosciences Division, Office of Basic Energy Sciences, Office
of Science, U.S. Department of Energy (Grant No. DE-FG02-01ER15228 to P.P)
and Phase I and II Software Fellowships awarded to J.E.D. by the
Molecular Sciences Software Institute funded by the National Science
Foundation grant ACI-1547580.
\end{acknowledgement}


\section*{APPENDIX A: PROOF OF THE EQUIVALENCE OF EQ \ref{A}, WITH THE CORRELATION
ENERGY DEFINED BY EQ \ref{C}, AND EQS \ref{fcc-singles}--\ref{fcc-energy} BASED ON EQS
\ref{t-def}, \ref{CC-property}, AND \ref{resolution}}

\renewcommand{\theequation}{A.\arabic{equation}}
\setcounter{equation}{0}

\setcounter{figure}{0}
\renewcommand{\thefigure}{A.\arabic{figure}}

\setcounter{chart}{0}
\renewcommand{\thechart}{A.\arabic{chart}}

The main purpose of this appendix is to demonstrate that the subsystem of CI equations
for the ground-state wave function $|\Psi\rangle$
defined by eqs \ref{eq-genciansatz}--\ref{eq-CPB}, corresponding to the projections on the singly and doubly excited
determinants, as in eq \ref{A}, where the correlation energy $\Delta E^\text{(CI)}$ is calculated using eq \ref{C},
can be transformed into the CC amplitude equations projected on singles and doubles, represented by eqs
\ref{cceq-singles} and \ref{cceq-doubles} or \ref{fcc-singles} and \ref{fcc-doubles},
with the CC energy given by eq \ref{fcc-energy}, if the cluster operator $T$ is defined by
eq \ref{t-def}. We begin by rewriting eq \ref{A} with the help of eq \ref{t-def}, which allows us to convert the CI
expansion for $|\Psi\rangle$, eq \ref{eq-genciansatz}, to a CC-type expression, eq \ref{eq-ccansatz}, while taking
advantage of the property of the exponential ansatz given by eq \ref{CC-property}.
We obtain,
\beq
\mel{\Phi_\alpha}{e^T (H_N e^T)_C}{\Phi} = \Delta E^\text{(CI)} \mel{\Phi_\alpha}{e^T}{\Phi},
\label{A1}
\eeq
where the CI correlation energy $\Delta E^\text{(CI)}$, eq \ref{C}, becomes
\beq
\Delta E^\text{(CI)} = \mel{\Phi}{H_N e^T}{\Phi}
= \mel{\Phi}{(H_N e^T)_{FC}}{\Phi},
\label{A2}
\eeq
in agreement with the CC energy formula given by eq \ref{fcc-energy}. Let us recall that the $\ket{\Phi_\alpha}$
states entering eq \ref{A1} are the determinants that span subspace $\mathscr{H}^{(P_\text{A})}$
matching the $C^{(P_\text{A})} \ket{\Phi}$ component of $|\Psi\rangle$ which, in this particular analysis,
are a shorthand notation for the singly and doubly excited determinants, $|\Phi_{i}^{a}\rangle$
and $|\Phi_{ij}^{ab}\rangle$, respectively.

The next step is the insertion of the resolution of the identity in the many-electron Hilbert space $\mathscr{H}$,
eq \ref{resolution}, between $e^T$ and $(H_N e^T)_C$ on the left-hand side of eq \ref{A1}. This allows us to
rewrite eq \ref{A1} as follows:
\beq
\Lambda_\alpha^{(1)} + \Lambda_\alpha^{(2)} + \Lambda_\alpha^{(3)} + 
\Lambda_\alpha^{(4)} = \Delta E^\text{(CI)} \mel{\Phi_\alpha}{e^T}{\Phi},
\label{A3}
\eeq
where
\beq
\Lambda_\alpha^{(1)} = \mel{\Phi_\alpha}{e^T}{\Phi} \mel{\Phi}{(H_N e^T)_{FC}}{\Phi}
\equiv \Delta E^\text{(CI)} \mel{\Phi_\alpha}{e^T}{\Phi},
\label{A4}
\eeq
\beq
\Lambda_\alpha^{(2)} = \sum_{\alpha^\prime} 
\mel{\Phi_\alpha}{e^T}{\Phi_{\alpha^\prime}} \mel{\Phi_{\alpha^\prime}}{(H_N e^T)_{C}}{\Phi} ,
\label{A5}
\eeq
\beq
\Lambda_\alpha^{(3)} = \sum_{\beta} \mel{\Phi_\alpha}{e^T}{\Phi_\beta} 
\mel{\Phi_\beta}{(H_N e^T)_{C}}{\Phi},
\label{A6}
\eeq
and
\beq
\Lambda_\alpha^{(4)} = \sum_{\gamma} \mel{\Phi_\alpha}{e^T}{\Phi_\gamma} 
\mel{\Phi_\gamma}{(H_N e^T)_{C}}{\Phi},
\label{A7}
\eeq
with $\ket{\Phi_\alpha},\ket{\Phi_{\alpha^\prime}} \in \mathscr{H}^{(P_\text{A})}$,
$\ket{\Phi_\beta} \in \mathscr{H}^{(P_\text{B})}$, and $\ket{\Phi_\gamma} \in \mathscr{H}^{(Q)}$.
We recall that $\mathscr{H}^{(P_\text{B})}$ is a subspace spanned by the determinants $\ket{\Phi_\beta}$
that match the content of the $C^{(P_\text{B})}$ operator, eq \ref{eq-CPB}, and
${\mathscr H}^{(Q)} = ({\mathscr H}^{(P)} \oplus \mathscr{H}^{(P_\text{A})} \oplus \mathscr{H}^{(P_\text{B})})^{\perp}$,
with ${\mathscr H}^{(P)}$ spanned by the reference determinant $|\Phi\rangle$,
is a subspace spanned by the remaining determinants, designated as $\ket{\Phi_\gamma}$,  which are not included
in the CI wave function $|\Psi\rangle$ defined by eqs \ref{eq-genciansatz}--\ref{eq-CPB}. In writing
eq \ref{A4}, we took advantage of the energy formula given by eq \ref{A2}.

Let us analyze each contribution $\Lambda_\alpha^{(k)}$, $k = 1\mbox{--}4$, to eq \ref{A3}. Since we have made
the assumption that the $\ket{\Phi_\alpha}$
states represent the singly and doubly excited determinants and $\ket{\Phi_\beta}$ and $\ket{\Phi_\gamma}$ are at least the triples,
the $\mel{\Phi_\alpha}{e^T}{\Phi_\beta}$ and $\mel{\Phi_\alpha}{e^T}{\Phi_\gamma}$ matrix elements in eqs \ref{A6} and \ref{A7}
vanish, i.e., $\Lambda_\alpha^{(3)} = \Lambda_\alpha^{(4)} = 0$. At the same time, $\Lambda_\alpha^{(1)}$, eq \ref{A4},
cancels out the right-hand side of eq \ref{A3}. This means that eq \ref{A3} reduces to
\beq
\Lambda_\alpha^{(2)} = 0 ,
\label{A8}
\eeq
with $\Lambda_\alpha^{(2)}$ defined by eq \ref{A5}. Since in the specific case considered here
$\ket{\Phi_\alpha}$ and $\ket{\Phi_{\alpha^\prime}}$ are the
singly and doubly excited determinants, the definition of the cluster operator $T$, eq \ref{t-def}, implies that
$e^{T} = 1 + C^{(P_\text{A})} + C^{(P_\text{B})}$, and the
$C_{2}$ and $C^{(P_\text{B})}$ operators generate at least double excitations, we can rewrite
the $\mel{\Phi_\alpha}{e^T}{\Phi_{\alpha^\prime}}$ term seen in eq \ref{A5} in the following manner:
\beq
\mel{\Phi_\alpha}{e^T}{\Phi_{\alpha^\prime}} \equiv \mel{\Phi_\alpha}{(1 + C_1
+ C_2 + C^{(P_\text{B})})}{\Phi_{\alpha^\prime}} = \delta_{\alpha \alpha^\prime}
+ \mel{\Phi_\alpha}{C_1}{\Phi_{\alpha^\prime}},
\label{A9}
\eeq
where $\delta_{\alpha \alpha^\prime}$ is the usual Kronecker delta. By substituting eq \ref{A9} into eq \ref{A5}
and using eq \ref{A8}, we immediately obtain
\beq
\Lambda_\alpha^{(2)} = \mel{\Phi_\alpha}{(H_Ne^T)_C}{\Phi}
+ \sum_{\alpha^\prime} \mel{\Phi_\alpha}{C_1}{\Phi_{\alpha^\prime}} \mel{\Phi_{\alpha^\prime}}{(H_Ne^T)_C}{\Phi}
= 0.
\label{A10}
\eeq

We now examine the system of equations represented by eq \ref{A10}. Let us start with the case in which
$\ket{\Phi_\alpha} = \ket{\Phi_i^a}$. Since $C_{1}$ generates single excitations and $\ket{\Phi_{\alpha^\prime}}$'s
in eq \ref{A10} are at least the singly excited determinants, $\mel{\Phi_\alpha}{C_1}{\Phi_{\alpha^\prime}} = 0$.
We can, therefore, conclude that the CI equations projected on the singly excited determinants, eq \ref{A}
with $\ket{\Phi_\alpha} = \ket{\Phi_i^a}$ or eq \ref{A-singles},
which are equivalent to eq \ref{A10} in which $\ket{\Phi_\alpha} = \ket{\Phi_i^a}$, reduce to the CC
equations projected on singles given by eq \ref{cceq-singles} or, more explicitly, eq \ref{fcc-singles}.

In the case of the projections on the doubly excited determinants $\ket{\Phi_{ij}^{ab}}$, i.e., when
$\ket{\Phi_\alpha} = \ket{\Phi_{ij}^{ab}}$, we can give eq \ref{A10} the following form:
\beq
\mel{\Phi_{ij}^{ab}}{(H_N e^T)_C}{\Phi} + \sum_{k,c} 
\mel{\Phi_{ij}^{ab}}{C_1}{\Phi_k^c} \mel{\Phi_k^c}{(H_N e^T)_C}{\Phi} = 0 ,
\label{A11}
\eeq
where we utilized the fact that the $\mel{\Phi_{ij}^{ab}}{C_1}{\Phi_{\alpha^{\prime}}}$ matrix element
vanishes unless $\ket{\Phi_{\alpha^{\prime}}}$ is a singly excited determinant. Since we have already
demonstrated that the cluster operator $T$, eq \ref{t-def}, satisfies eq \ref{cceq-singles},
i.e., $\mel{\Phi_k^c}{(H_N e^T)_C}{\Phi} = 0$, eq \ref{A11} simplifies to eq \ref{cceq-doubles}.
This means that the CI equations projected on the doubly excited determinants, eq \ref{A} with
$\ket{\Phi_\alpha} = \ket{\Phi_{ij}^{ab}}$ or eq \ref{A-doubles},
which are equivalent to eq \ref{A10} in which $\ket{\Phi_\alpha} = \ket{\Phi_{ij}^{ab}}$, reduce to the CC
equations projected on doubles given by eq \ref{cceq-doubles} or, more explicitly, eq \ref{fcc-doubles}.
This concludes our first proof of the equivalence of the CI eqs \ref{A} and \ref{C} and their CC
counterparts represented by eqs \ref{cceq-singles}, \ref{cceq-doubles}, and \ref{fcc-energy} or
\ref{fcc-singles}--\ref{fcc-energy}.

As mentioned at the end of Section \ref{section2.1}, one can extend the above considerations
to higher-order ec-CC variants, in which the excitation operators $C^{(P_\text{A})}$ and
$C^{(P_\text{B})}$ that enter the CI wave function $|\Psi \rangle$ through eq \ref{eq-genciansatz}
are defined by eqs \ref{eq-CPA-ex} and \ref{eq-CPB-ex}. In this case, subspace
$\mathscr{H}^{(P_\text{A})}$, which matches the content of $C^{(P_\text{A})}$, is spanned
by all determinants $|\Phi_{\alpha} \rangle$ with the excitation ranks ranging from 1 to $m_{A}$,
where $m_{A} \geq 2$,
and determinants $|\Phi_{\beta} \rangle \in \mathscr{H}^{(P_\text{B})}$ that match the
many-body components $C_n^{(P_\text{B})}$ of operator $C^{(P_\text{B})}$, assuming that
$C^{(P_\text{B})} \neq 0$, have the excitation
ranks exceeding $m_{A}$. In order to prove the equivalence of eq \ref{A}, with
the correlation energy defined by \ref{C}, and the CC system defined by eq \ref{eq-CCA}
in this generalized case, we follow the same procedure as described above, adjusting it to
the contents of operators $C^{(P_\text{A})}$ and $C^{(P_\text{B})}$ and subspaces
$\mathscr{H}^{(P_\text{A})}$ and $\mathscr{H}^{(P_\text{B})}$. It is immediately obvious that
eqs \ref{A1}--\ref{A8} still hold. In particular, $\Lambda_\alpha^{(1)}$ cancels out
the right-hand side of eq \ref{A3} and matrix elements
$\mel{\Phi_\alpha}{e^T}{\Phi_\beta}$ and $\mel{\Phi_\alpha}{e^T}{\Phi_\gamma}$ that enter the
$\Lambda_\alpha^{(3)}$ and $\Lambda_\alpha^{(4)}$ contributions to eq \ref{A3} vanish, since the many-body
ranks of determinants $|\Phi_{\alpha} \rangle \in \mathscr{H}^{(P_\text{A})}$ do not exceed $m_{A}$,
the excitation levels of $|\Phi_{\beta} \rangle$ and $|\Phi_{\gamma} \rangle$, which belong to
$\mathscr{H}^{(P_\text{B})}$ and $\mathscr{H}^{(Q)}$, respectively, are at least
$m_{A}+1$, and $e^{T} = 1 + C^{(P_\text{A})} + C^{(P_\text{B})}$. With the generalized
definitions of the excitation operators $C^{(P_\text{A})}$ and $C^{(P_\text{B})}$
considered here, eq \ref{A8} can be given the following form:
\beq
\Lambda_\alpha^{(2)} = \mel{\Phi_\alpha}{(H_Ne^T)_C}{\Phi}
+ \sum_{\alpha^\prime} \mel{\Phi_\alpha}{\widetilde{C}}{\Phi_{\alpha^\prime}}
\mel{\Phi_{\alpha^\prime}}{(H_Ne^T)_C}{\Phi} = 0 ,
\label{A12}
\eeq
where
\beq
\widetilde{C} = \sum_{n=1}^{m_{A}-1} C_{n} ,
\label{A13}
\eeq
since the excitation ranks of determinants $|\Phi_{\alpha} \rangle$ and
$|\Phi_{\alpha^{\prime}} \rangle$ belonging to $\mathscr{H}^{(P_\text{A})}$ range from 1 to $m_{A}$
and operator $C^{(P_\text{B})}$ generates higher--than--$m_{A}$-fold excitations, so that
\beq
\mel{\Phi_\alpha}{e^T}{\Phi_{\alpha^\prime}} \equiv \mel{\Phi_\alpha}{(1 + \sum_{n=1}^{m_{A}} C_{n}
+ C^{(P_\text{B})})}{\Phi_{\alpha^\prime}} = \delta_{\alpha \alpha^\prime}
+ \mel{\Phi_\alpha}{\widetilde{C}}{\Phi_{\alpha^\prime}} ,
\label{A14}
\eeq
with $\widetilde{C}$ defined by eq \ref{A13}.

As in the previously considered $m_{A} = 2$ case, we examine the system represented by eq \ref{A12},
which we can do in a recursive manner starting from $\ket{\Phi_\alpha} = \ket{\Phi_i^a}$.
When $\ket{\Phi_\alpha} = \ket{\Phi_i^a}$, matrix element $\mel{\Phi_\alpha}{\widetilde{C}}{\Phi_{\alpha^\prime}}$
vanishes, since $|\Phi_{\alpha^{\prime}} \rangle$ is at least a singly excited determinant and
$\widetilde{C}$ defined by eq \ref{A13} generates at least single excitations. This immediately leads to the CC equations
projected on the singly excited determinants, eq \ref{cceq-singles}, i.e., eq \ref{eq-CCA} is
satisfied when $\ket{\Phi_\alpha} = \ket{\Phi_i^a}$. Moving on, when $\ket{\Phi_\alpha} = \ket{\Phi_{ij}^{ab}}$,
$\mel{\Phi_\alpha}{\widetilde{C}}{\Phi_{\alpha^\prime}} = 0$ unless $|\Phi_{\alpha^{\prime}} \rangle$
is a singly excited determinant, but since the CC equations projected on singles are already satisfied,
the summation over $\alpha^{\prime}$ in eq \ref{A12} vanishes and eq \ref{A12} reduces to the
CC equations projected on doubles, eq \ref{cceq-doubles}, which means that eq \ref{eq-CCA} remains
true for $\ket{\Phi_\alpha} = \ket{\Phi_{ij}^{ab}}$. Continuing (assuming that $m_{A} > 2$),
when $\ket{\Phi_\alpha} = \ket{\Phi_{ijk}^{abc}}$, $\mel{\Phi_\alpha}{\widetilde{C}}{\Phi_{\alpha^\prime}}$
is zero unless $|\Phi_{\alpha^{\prime}} \rangle$ is a singly or doubly excited determinant.
Again, since the CC equations projected on singles and doubles are already satisfied,
the summation over $\alpha^{\prime}$ in eq \ref{A12} becomes zero and
we obtain $\mel{\Phi_{ijk}^{abc}}{(H_N e^{T})_{C}}{\Phi} = 0$, i.e., eq \ref{eq-CCA} is
satisfied in the $\ket{\Phi_\alpha} = \ket{\Phi_{ijk}^{abc}}$ case. One can repeat the same procedure
for the projections on higher-rank determinants $\ket{\Phi_\alpha}$ in eq \ref{A12} that belong to subspace
$\mathscr{H}^{(P_\text{A})}$. In the final stage of this recursive analysis, i.e., when
$\ket{\Phi_\alpha}$ is an $m_{A}$-tuply excited determinant, eq \ref{eq-CCA} remains true as well,
since $\mel{\Phi_\alpha}{\widetilde{C}}{\Phi_{\alpha^\prime}}$ vanishes unless the excitation
rank of $|\Phi_{\alpha^{\prime}} \rangle$ is at most $m_{A}-1$ and the CC equations projected
on up to $(m_{A}-1)$-tuply excited determinants are already satisfied. This means that
$\mel{\Phi_\alpha}{(H_N e^{T})_{C}}{\Phi}$ is zero, i.e., eq \ref{eq-CCA} holds once again. In other
words, eq \ref{eq-CCA} remains true for all determinants $|\Phi_{\alpha} \rangle \in \mathscr{H}^{(P_\text{A})}$,
i.e., eq \ref{A}, with the correlation energy defined by eq \ref{C}, is equivalent to the CC system defined
by eq \ref{eq-CCA} when the excitation operators $C^{(P_\text{A})}$ and
$C^{(P_\text{B})}$ that enter the CI wave function $|\Psi \rangle$ through eq \ref{eq-genciansatz}
are defined by eqs \ref{eq-CPA-ex} and \ref{eq-CPB-ex} and subspace
$\mathscr{H}^{(P_\text{A})}$, which matches the content of $C^{(P_\text{A})}$, is spanned
by all determinants $|\Phi_{\alpha} \rangle$ with the excitation ranks ranging from 1 to $m_{A}$.

\section*{APPENDIX B: DIAGRAMMATIC PROOF OF THE EQUIVALENCE OF EQ \ref{A}, WITH THE CORRELATION
ENERGY DEFINED BY EQ \ref{C}, OR EQS \ref{A-singles}--\ref{C-alt} AND EQS  \ref{fcc-singles}--\ref{fcc-energy}}

\renewcommand{\theequation}{B.\arabic{equation}}
\setcounter{equation}{0}

\setcounter{figure}{0}
\renewcommand{\thefigure}{B.\arabic{figure}}

\setcounter{chart}{0}
\renewcommand{\thechart}{B.\arabic{chart}}

In this appendix, we provide an alternative proof of the equivalence of eqs \ref{fcc-singles}--\ref{fcc-energy}
and \ref{A-singles}--\ref{C-alt} using a diagrammatic approach.
As in the case of the algebraic derivation presented in Appendix A, the only assumption
that we make regarding the CI state $|\Psi\rangle$ used to provide information about the
$T_{3}$ and $T_{4}$ clusters for ec-CC considerations is the full treatment of the $C_1$ and $C_2$ components.
This means that all of the mathematical manipulations in this appendix apply to conventional as well as
unconventional truncations in the CI excitation operator, as defined by eqs \ref{eq-genciansatz}--\ref{eq-CPB},
in addition to FCI.
The diagrammatic derivation of the equivalence of eqs \ref{fcc-singles}--\ref{fcc-energy} and
\ref{A-singles}--\ref{C-alt} is accomplished by starting from the CC equations corresponding to
projections on singles and doubles, eqs \ref{fcc-singles} and \ref{fcc-doubles}, respectively, and, after
performing cluster analysis of the CI wave function $|\Psi\rangle$ with the help of eq \ref{cluster_analysis}, converting
them to the analogous eqs \ref{A-singles} and \ref{A-doubles}, with the correlation energy defined
by eq \ref{C-alt}, which are part of the CI eigenvalue problem for $|\Psi\rangle$ used to
determine the three- and four-body clusters.
To facilitate our presentation, throughout this appendix we drop the
`$(P_\text{B})$' superscript in the $C_n^{(P_\text{B})}$ components of operator $C^{(P_\text{B})}$,
eq \ref{eq-CPB}, associated with higher--than--doubly excited contributions to $|\Psi\rangle$.

The first step is to express eqs \ref{fcc-singles} and \ref{fcc-doubles} in terms of the $C_1 \text{--} C_4$
operators by using the relationships between the $T_{n}$ and $C_{n}$ components given by eq \ref{cluster_analysis}.
In the case of the singles projections, the correspondence between the various $(H_N e^T)_C$ terms appearing
in eq \ref{fcc-singles} and their counterparts resulting from the application of eq \ref{cluster_analysis}
is provided in Chart \ref{chart-B1}. As shown in Chart \ref{chart-B1}, all contributions containing
products of $C_{n}$ components other than the unlinked terms, marked in red, in which the fully connected
operator products $(F_N \mbf{C_1})_{FC}$ and $(V_N \mbf{C_2})_{FC}$ that contribute to the correlation
energy multiply $C_1$, cancel out. As a result, the CC equations
projected
on the singly excited determinants, eq \ref{fcc-singles}, become
\beq
\mel{\Phi_i^a}{[F_N + (F_N \mbf{C_1})_C + (F_N \mbf{C_2})_C + (V_N \mbf{C_1})_C + (V_N \mbf{C_2})_C
	+ (V_N \mbf{C_3})_C + \Theta_1 ]} {\Phi} = 0,
\label{B1}
\eeq
where
\beq
\Theta_1 = -[(F_N \mbf{C_1})_{FC} \mbf{C_1} + (V_N \mbf{C_2})_{FC} \mbf{C_1}]
\label{B2}
\eeq
represents the terms highlighted in Chart \ref{chart-B1} in red. 
Focusing on
$\Theta_1$,
we obtain
\beq
\begin{split}
	\mel{\Phi_i^a}{\Theta_1}{\Phi} &= -\mel{\Phi_i^a} {[(F_N \mbf{C_1})_{FC} \mbf{C_1} + (V_N
		\mbf{C_2})_{FC} \mbf{C_1}]}{\Phi} \\
	&= -\mel{\Phi_i^a} {[(F_N \mbf{C_1})_{FC} + (V_N \mbf{C_2})_{FC}] \mbf{C_1}} {\Phi} \\
	&= -\Delta E^\text{(CI)} \mel{\Phi_i^a} {\mbf{C_1}} {\Phi},
\label{B3}
\end{split}
\eeq
where we used eq \ref{C-alt} for the CI correlation energy $\Delta E^\text{(CI)}$, which, after replacing
$H_N$ by the sum of $F_N$ and $V_N$, is equivalent to
\beq
\Delta E^{(\text{CI})} = \mel{\Phi} {[(F_N C_1)_{FC} + (V_N C_2)_{FC}]} {\Phi} .
\label{B4}
\eeq
Inserting eq \ref{B3} for $\mel{\Phi_i^a}{\Theta_1}{\Phi}$ back to eq \ref{B1} and
moving the energy-dependent term $\Delta E^\text{(CI)} \mel{\Phi_i^a} {\mbf{C_1}} {\Phi}$
to the right-hand side of the resulting expression, we arrive at
\beq
\mel{\Phi_i^a} {[F_N + (F_N \mbf{C_1})_C + (F_N \mbf{C_2})_C +
(V_N \mbf{C_1})_C + (V_N \mbf{C_2})_C + (V_N \mbf{C_3})_C]}{\Phi}
= \Delta E^{(\text{CI})} \mel{\Phi_i^a}{\mbf{C_1}} {\Phi},
\label{B5}
\eeq
which is equivalent to eq \ref{A-singles}, when expressed in terms of
the one- and two-body components of $H_N$. This completes the proof of the equivalence of eqs
\ref{fcc-singles} and \ref{A-singles}.

\begin{figure*}
\centering
\captionsetup[subfigure]{labelfont={large,bf}}
\subfloat[]{\includegraphics[width=0.3\textwidth]{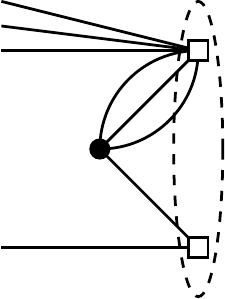}}\hfill
\subfloat[]{\includegraphics[width=0.3\textwidth]{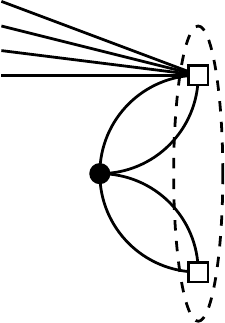}}\hfill
\subfloat[]{\includegraphics[width=0.3\textwidth]{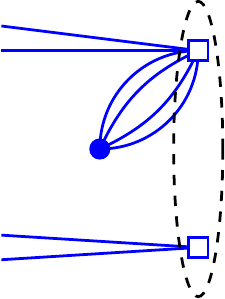}}\\
\subfloat[]{\includegraphics[width=0.3\textwidth]{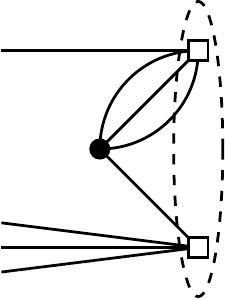}}\hfill
\subfloat[]{\includegraphics[width=0.3\textwidth]{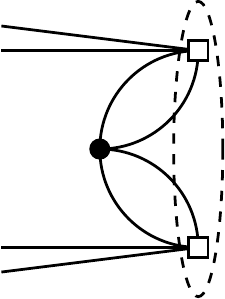}}\hfill
\subfloat[]{\includegraphics[width=0.3\textwidth]{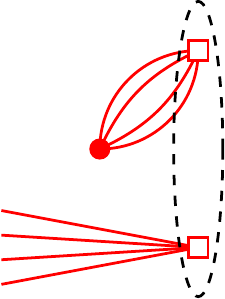}}\\
\vspace*{12pt}
\caption{\label{figure-B1}
Hugenholtz diagrammatic representation of the $V_N C_1 C_3$ ((a)--(c)) and $\tfrac{1}{2} V_N C_2^2$
((d)--(f)) contributions to $\mel{\Phi_{ij}^{ab}}{(V_N T_4)_C}{\Phi}$ resulting from the
application of eq \ref{cluster_analysis} and corresponding to the second through fifth expressions
contributing to $(V_N T_4)_C$ in Chart \ref{chart-B2}. The solid circle represents the $V_N$ vertex and
the open squares with two, four, and six fermion lines designate the $C_1$, $C_2$, and
$C_3$ operators, respectively. The dashed oval highlights the $T_4$ operator in
$\mel{\Phi_{ij}^{ab}}{(V_N T_4)_C}{\Phi}$. Consistent with Chart \ref{chart-B2},
diagram (c), which corresponds to the disconnected $(V_N C_3)_{C} C_1$ contribution
to $\mel{\Phi_{ij}^{ab}}{(V_N T_4)_C}{\Phi}$ (the third term contributing to $(V_N T_4)_C$ in Chart \ref{chart-B2})
is highlighted in blue. Diagram (f), which represents the unlinked $(V_N C_2)_{FC} C_2$
contribution to $\mel{\Phi_{ij}^{ab}}{(V_N T_4)_C}{\Phi}$ (the fifth term contributing to
$(V_N T_4)_C$ in Chart \ref{chart-B2}) is highlighted in red.
}
\end{figure*}

A similar analysis can be performed for the CC equations corresponding to projections on the doubly excited
determinants, eq \ref{fcc-doubles}. The correspondence between the various $(H_N e^T)_C$ terms contributing
to eq \ref{fcc-doubles} and their counterparts obtained by using eq \ref{cluster_analysis} is shown
in Chart \ref{chart-B2}. In this case, despite the cancellation of the majority of the
nonlinear terms in the $C_{n}$ components resulting from the application of eq \ref{cluster_analysis}, the
emergence of the CI equations projected on doubles, eq \ref{A-doubles}, from eq \ref{fcc-doubles} is not as obvious
as in the previously examined singles projections. As shown in Chart \ref{chart-B2}, in addition to the unlinked
terms, marked in red and green, in which the fully connected operator products
$(F_N \mbf{C_1})_{FC}$ and $(V_N \mbf{C_2})_{FC}$
that contribute to the correlation energy multiply $C_2$ and $\tfrac{1}{2} C_1^2$, we see the appearance
of the disconnected quantities, marked in blue, where the $(F_N \mbf{C_1})_C$,
$(F_N \mbf{C_2})_C$, $(V_N \mbf{C_1})_C$, $(V_N \mbf{C_2})_C$, and $(V_N \mbf{C_3})_C$
connected operator products multiply $C_1$. The Hugenholtz diagrams
emerging from the $V_N C_1 C_3$ and $\tfrac{1}{2} V_N C_2^2$ operator products,
which result from the application of eq \ref{cluster_analysis} to the $\mel{\Phi_{ij}^{ab}}{(V_N T_4)_C}{\Phi}$
contribution to eq \ref{fcc-doubles} and which correspond to the second through fifth expressions
contributing to $(V_N T_4)_C$ in Chart \ref{chart-B2}, are shown in Figure \ref{figure-B1}. It should be noted
that the $T_4$ component (emphasized in Figure \ref{figure-B1} by a dashed oval)
resulting from the cluster analysis defined by eq \ref{cluster_analysis}
is a strictly connected quantity (in an MBPT sense) only in a FCI limit.

After removing the nonlinear terms in $C_{n}$ components that cancel out and grouping the remaining contributions
to the CC equations corresponding to projections on the doubly excited determinants according to their color
in Chart \ref{chart-B2}, eq \ref{fcc-doubles} becomes
\beq
\begin{split}
	\bra{\Phi_{ij}^{ab}}[(F_N \mbf{C_2})_C &+ (F_N \mbf{C_3})_C + V_N + (V_N \mbf{C_1})_C 
	+ (V_N \mbf{C_2})_C + (V_N \mbf{C_3})_C + (V_N \mbf{C_4})_C \\
	& + \Theta^\prime  + \Theta^{\prime \prime} + 
	\Theta_2] \ket{\Phi} = 0,
\end{split}
\label{B6}
\eeq
where
\beq
\Theta^\prime = -[(F_N \mbf{C_1})_C \mbf{C_1} + (F_N \mbf{C_2})_C \mbf{C_1} + (V_N \mbf{C_1})_C \mbf{C_1}
+ (V_N \mbf{C_2})_C \mbf{C_1} + (V_N \mbf{C_3})_C \mbf{C_1}],
\label{B7}
\eeq
\beq
\Theta^{\prime \prime} = 2(F_N \mbf{C_1})_{FC} \tfrac{1}{2} \mbf{C_1}^2
+ 2 (V_N \mbf{C_2})_{FC} \tfrac{1}{2} \mbf{C_1}^2,
\label{B8}
\eeq
and
\beq
\Theta_2 = -[(F_N \mbf{C_1})_{FC} \mbf{C_2} + (V_N \mbf{C_2})_{FC} \mbf{C_2}] .
\label{B9}
\eeq
After adding and subtracting $F_N \mbf{C_1}$ on the left-hand side of eq \ref{B6}, we obtain
\beq
\begin{split}
	\bra{\Phi_{ij}^{ab}} \![F_N \mbf{C_1} &+ (F_N \mbf{C_2})_C + (F_N \mbf{C_3})_C + V_N 
	+ (V_N \mbf{C_1})_C + (V_N \mbf{C_2})_C + (V_N \mbf{C_3})_C \\
	& + (V_N \mbf{C_4})_C + \widetilde{\Theta}^{\prime}  + \Theta^{\prime \prime} + \Theta_2]\! \ket{\Phi} = 0,
\end{split}
\label{B10}
\eeq
where
\beq
\widetilde{\Theta}^{\prime} = \Theta^{\prime} - F_N \mbf{C_1}.
\label{B11}
\eeq
Using eq \ref{B7}, factoring out $C_1$, and taking advantage of the already obtained CI equations projected
on the singly excited determinants, eq \ref{B5}, the contribution from the $\widetilde{\Theta}^\prime$ term,
defined by eq \ref{B11}, to eq \ref{B10}, can be rewritten as follows:
\beq
\begin{split}
	\mel{\Phi_{ij}^{ab}}{\widetilde{\Theta}^\prime}{\Phi} &=
	-\bra{\Phi_{ij}^{ab}} \![F_N \mbf{C_1} + (F_N \mbf{C_1})_C \mbf{C_1} + (F_N \mbf{C_2})_C 
	\mbf{C_1} + (V_N \mbf{C_1})_C \mbf{C_1} \\
	&\phantom{= -\bra{\Phi_{ij}^{ab}}[} + (V_N \mbf{C_2})_C \mbf{C_1} + (V_N \mbf{C_3})_C \mbf{C_1}]\! \ket{\Phi} \\
        &= -\bra{\Phi_{ij}^{ab}} \![F_N + (F_N \mbf{C_1})_C + (F_N \mbf{C_2})_C + (V_N \mbf{C_1})_C + (V_N \mbf{C_2})_C \\
        &\phantom{= -\bra{\Phi_{ij}^{ab}}[} + (V_N \mbf{C_3})_C] \mbf{C_1}\! \ket{\Phi} \\
	&= -\Delta E^\text{(CI)} \mel{\Phi_{ij}^{ab}}{\mbf{C_1}^2}{\Phi} .
\end{split}
\label{B12}
\eeq	
At the same time, the contributions from the unlinked
$\Theta^{\prime \prime}$ and $\Theta_2$ terms to eq \ref{B10}, after factoring out
$C_{1}^{2}$ and $C_{2}$, respectively, and using eq \ref{B4} for the CI correlation energy, become
\beq
\begin{split}
	\mel{\Phi_{ij}^{ab}}{\Theta^{\prime \prime}}{\Phi} &= \mel{\Phi_{ij}^{ab}}{[2(F_N 
		\mbf{C_1})_{FC} \tfrac{1}{2} \mbf{C_1}^2 + 2 (V_N \mbf{C_2})_{FC} \tfrac{1}{2} \mbf{C_1}^2]}{\Phi} \\
	&= \mel{\Phi_{ij}^{ab}}{[(F_N \mbf{C_1})_{FC} + (V_N \mbf{C_2})_{FC}] \mbf{C_1}^2}{\Phi} \\
	&= \Delta E^\text{(CI)} \mel{\Phi_{ij}^{ab}}{\mbf{C_1}^2}{\Phi}
\end{split}
\label{B13}
\eeq
and
\beq
\begin{split}
	\mel{\Phi_{ij}^{ab}}{\Theta_2}{\Phi} &= - \mel{\Phi_{ij}^{ab}}{[(F_N \mbf{C_1})_{FC} \mbf{C_2}
		+ (V_N \mbf{C_2})_{FC} \mbf{C_2}]}{\Phi} \\
	&= -\mel{\Phi_{ij}^{ab}}{[(F_N \mbf{C_1})_{FC} + (V_N \mbf{C_2})_{FC}] \mbf{C_2}}{\Phi} \\
	&= -\Delta E^{(\text{CI})} \mel{\Phi_{ij}^{ab}}{\mbf{C_2}}{\Phi} .
\label{B14}
\end{split}
\eeq
Note that after all of these manipulations the
$\widetilde{\Theta}^\prime$ and $\Theta^{\prime \prime}$ contributions to eq \ref{B10},
eqs \ref{B12} and \ref{B13}, respectively, cancel each other. Thus, after inserting
eq \ref{B14} back to eq \ref{B10} and moving the energy-dependent
$\Delta E^{(\text{CI})} \mel{\Phi_{ij}^{ab}}{\mbf{C_2}}{\Phi}$ contribution to the
right-hand side of the resulting formula, we arrive at
\beq
        \begin{split}
        \bra{\Phi_{ij}^{ab}} \![F_N \mbf{C_1} &+ (F_N
        \mbf{C_2})_C + (F_N \mbf{C_3})_C + V_N + (V_N \mbf{C_1})_C + (V_N
        \mbf{C_2})_C \\
        &+
        (V_N \mbf{C_3})_C +
        (V_N \mbf{C_4})_C]\! \ket{\Phi} = \Delta E^{(\text{CI})}
        \mel{\Phi_{ij}^{ab}} {\mbf{C_2}} {\Phi},
        \end{split}
\label{B15}
\eeq
which is equivalent to eq \ref{A-doubles}, when expressed in terms of
the one- and two-body components of $H_N$. This completes the proof of the equivalence of eqs
\ref{fcc-doubles} and \ref{A-doubles} and the diagrammatic derivation of the CI
eqs  \ref{A-singles}--\ref{C-alt} from the CC eqs \ref{fcc-singles}--\ref{fcc-energy}.
The equivalence of eqs \ref{fcc-energy} and \ref{C-alt} for the correlation energy
is an obvious consequence of replacing $T_{1}$ and $T_{2}$ in eq \ref{fcc-energy} by
the formulas in terms of $C_{1}$ and $C_{2}$ originating from eq \ref{cluster_analysis}, which
leads directly to eq \ref{B4} and its analog, eq \ref{C-alt}.


\newpage


\providecommand{\latin}[1]{#1}
\makeatletter
\providecommand{\doi}
  {\begingroup\let\do\@makeother\dospecials
  \catcode`\{=1 \catcode`\}=2 \doi@aux}
\providecommand{\doi@aux}[1]{\endgroup\texttt{#1}}
\makeatother
\providecommand*\mcitethebibliography{\thebibliography}
\csname @ifundefined\endcsname{endmcitethebibliography}
  {\let\endmcitethebibliography\endthebibliography}{}


\pagebreak

\begin{table*}
	\caption{\label{table1} A comparison of the energies resulting from the various 
	CI and CC all-electron calculations for the H$_2$O molecule, as described by the 
	cc-pVDZ basis set,\cite{Dunning1989} at the equilibrium and two displaced 
	geometries in which both O--H bonds are stretched by factors of 2 and 
	3.\textsuperscript{\emph{a}}}
	\begin{tabular}{l d d d}
		\hline\hline
		\multirow{2}{*}{wave function} & 
		\multicolumn{1}{c}{\multirow{2}{*}{CI/CC energy}} & \multicolumn{2}{c}{ec-CC 
		energy}\\
		\cline{3-4}
		& & \multicolumn{1}{c}{I} & \multicolumn{1}{c}{II} \\
		\hline
		\multicolumn{4}{c}{\underline{$R = R_{e}$}} \\
		CISD     & 12.023 & 12.023 & 3.744\textsuperscript{\emph{b}} \\
		CISDT    &  9.043 &  9.043 & 0.455 \\
		CISDTQ   &  0.327 &  0.327 & 0.327 \\
		CISDTQP  &  0.139 &  0.139 & 0.139 \\
		CISDTQPH &  0.003 &  0.003 & 0.003 \\
		CCSD\textsuperscript{\emph{c}} & 3.744 & 3.744 & 3.744 \\
		CCSDT\textsuperscript{\emph{c}} & 0.493 & 0.493 & 0.493 \\
		CCSDTQ\textsuperscript{\emph{c}} & 0.019 & 0.019 & 0.019 \\
		FCI\textsuperscript{\emph{d}} & -76.241860 & & \\

		\multicolumn{4}{c}{\underline{$R = 2 R_{e}$}} \\
		CISD     & 72.017 & 72.017 & 22.034\textsuperscript{\emph{b}} \\
		CISDT    & 56.096 & 56.096 &  2.920 \\
		CISDTQ   &  5.819 &  5.819 &  5.819 \\
		CISDTQP  &  2.236 &  2.236 &  2.236 \\
		CISDTQPH &  0.059 &  0.059 &  0.059 \\
		CCSD\textsuperscript{\emph{c}}   & 22.034 & 22.034 & 22.034 \\
		CCSDT\textsuperscript{\emph{c}}  & -1.403 & -1.403 & -1.403 \\
		CCSDTQ\textsuperscript{\emph{c}} &  0.032 &  0.032 &  0.032 \\
		FCI\textsuperscript{\emph{d}} & -75.951667 & & \\
		\multicolumn{4}{c}{\underline{$R = 3R_{e}$}} \\
		CISD     & 164.949 & 164.949 &  10.849\textsuperscript{\emph{b}} \\
		CISDT    & 118.119 & 118.119 & -77.317 \\
		CISDTQ   &  16.150 &  16.150 &  16.150 \\
		CISDTQP  &   6.432 &   6.432 &   6.432 \\
		CISDTQPH &   0.159 &   0.159 &   0.159 \\
		CCSD\textsuperscript{\emph{c}}  & 10.849 & 10.849 & 10.849 \\
		CCSDT\textsuperscript{\emph{c}} & -40.126 & -40.126 & -40.126 \\
		CCSDTQ\textsuperscript{\emph{c}} & -4.733 & -4.733 & -4.733 \\
		FCI\textsuperscript{\emph{d}} & -75.911946 & & \\
		
		\hline\hline
	\end{tabular}
	\vspace*{1mm}

	 {\footnotesize
	 \textsuperscript{\emph{a}} The equilibrium geometry, $R = R_{e}$, and the 
	 geometries that represent a simultaneous stretching of both O--H bonds by 
	 factors of 2 and 3 without changing the $\angle\text{(H--O--H)}$ angle were 
	 taken from ref \citenum{olsen-h2o}. Unless otherwise stated, all energies are 
	 errors relative to FCI in millihartree.
	 \textsuperscript{\emph{b}} Equivalent to CCSD.
	 \textsuperscript{\emph{c}} Taken from ref \citenum{Bauman2017}.
	 \textsuperscript{\emph{d}} Total FCI energy in hartree.
	}
\end{table*}

\pagebreak

\begin{landscape}
\begin{table*}
	\renewcommand{\arraystretch}{0.68}
	\caption{\label{table2} Convergence of the energies resulting from the 
	all-electron CIPSI 
	calculations initiated from the RHF wave function and the corresponding 
	CIPSI-based ec-CC energies toward FCI for the H$_2$O molecule, as described by 
	the cc-pVDZ basis set,\cite{Dunning1989} at the equilibrium and two displaced 
	geometries in which both O--H bonds are stretched by factors of 2 and 
	3.\textsuperscript{\emph{a}}}		
\begin{tabular}{c @{\extracolsep{18pt}} r @{\extracolsep{16pt}} r @{\extracolsep{16pt}}
r @{\extracolsep{16pt}} r @{\extracolsep{18pt}} r @{\extracolsep{4pt}} r @{\extracolsep{4pt}} r @{\extracolsep{18pt}}
r @{\extracolsep{10pt}} r @{\extracolsep{10pt}}r}
			\hline\hline
			\multirow{2}{*}{$N_\text{det(in)}$ / $N_\text{det(out)}$} & 
			\multirow{2}{*}{\%S\textsuperscript{\emph{b}}} & 
			\multirow{2}{*}{\%D\textsuperscript{\emph{b}}} & 
			\multirow{2}{*}{\%T\textsuperscript{\emph{b}}} & 
			\multirow{2}{*}{\%Q\textsuperscript{\emph{b}}} &
			\multicolumn{3}{c}{CIPSI\textsuperscript{\emph{c}}}  &
			\multicolumn{3}{c}{ec-CC\textsuperscript{\emph{c}}} \\
			\cline{6-8} \cline{9-11} 
			&  &  & &  & \multicolumn{1}{c}{$E_\text{var}$} & 
			$E_\text{var} + \Delta E^{(2)}$ & 
			$E_\text{var} + \Delta E_{r}^{(2)}$ &
			\multicolumn{1}{c}{I} & \multicolumn{1}{c}{II} & 
			\multicolumn{1}{c}{II$_3$} \\
			\hline
			\multicolumn{11}{c}{\underline{$R = R_{e}$}} \\
			    1 / 1 & 0 & 0 & 0 & 0 & 217.822\textsuperscript{\emph{d}} & 
			    $-42.098$ & $-21.684$ &
			    3.744\textsuperscript{\emph{e}} & 3.744\textsuperscript{\emph{e}} & 
			    0.344\textsuperscript{\emph{f}} \\
			1,000 / 1,299 & 15.2 & 37.9  & 0 & 0.0 & 21.589 & $-0.109$ & $-0.024$ & 11.019 & 
			3.637 & 0.236 \\
			5,000 / 5,216 & 51.5 & 73.8 & 1.0 & 0.2 & 8.445 & 0.098 & 0.108 & 8.123 & 2.455 
			& 0.012 \\
			10,000 / 10,448 & 60.6 & 80.4 & 3.1 & 0.4 & 6.587 & 0.184 & 0.190 & 6.441 & 
			1.887 & 0.063 \\
			50,000 / 83,762 & 93.9 & 94.7 & 22.1 & 4.9 & 2.612 & 0.089 & 0.090 & 2.610 & 
			0.626 & 0.168 \\
			100,000 / 167,425 & 93.9 & 97.8 & 34.5 & 10.0 & 1.743 & 0.064 & 0.064 & 1.742 & 
			0.410 & 0.190 \\
			500,000 / 665,840 & 100 & 99.7 & 69.0 & 31.9 & 0.435 & 0.008 & 0.008 & 0.435 & 
			0.120 & 0.105 \\
			1,000,000 / 1,308,003 & 100 & 99.9 & 82.5 & 45.6 & 0.229 & 0.004 & 0.004 & 0.229 
			& 0.069 & 0.066 \\
			\multicolumn{11}{c}{\underline{$R = 2 R_{e}$}} \\
			1 / 1 & 0 & 0 & 0 & 0 & 363.956\textsuperscript{\emph{d}} & $-180.621$ & 105.169 &
			22.034\textsuperscript{\emph{e}} & 22.034\textsuperscript{\emph{e}} & 
			$-0.548$\textsuperscript{\emph{f}} \\
			1,000 / 1,399 & 36.4 & 18.1 & 0.3 & 0.0 & 34.996 & 1.884 & 2.139 & 26.505 & 
			9.639 & $-0.454$ \\
			5,000 / 5,664 & 54.5 & 43.1 & 1.8 & 0.2 & 13.817 & 0.605 & 0.638 & 12.707 & 
			4.102 & 0.226 \\
			10,000 / 11,350 & 57.6 & 55.6 & 3.6 & 0.4 & 9.011 & 0.375 & 0.388 & 8.653 & 
			2.677 & 0.321 \\
			50,000 / 90,880 & 93.9 & 85.3 & 21.6 & 3.7 & 2.436 & 0.084 & 0.085 & 2.429 & 
			0.788 & 0.515 \\
			100,000 / 181,579 & 100 & 91.6 & 31.7 & 6.8 & 1.418 & 0.046 & 0.046 & 1.417 & 
			0.467 & 0.356 \\
			500,000 / 718,316 & 100 & 97.7 & 59.6 & 21.4 & 0.273 & 0.009 & 0.009 & 0.273 & 
			0.147 & 0.138 \\
			1,000,000 / 1,390,678 & 100 & 99.3 & 73.2 & 31.9 & 0.137 & 0.003 & 0.003 & 0.137 
			& 0.074 & 0.072 \\
			\multicolumn{11}{c}{\underline{$R = 3 R_{e}$}} \\
			1 / 1 & 0 & 0 & 0 & 0 & 567.554\textsuperscript{\emph{d}} & $-227.583$ & 420.893 &
			10.849\textsuperscript{\emph{e}} & 10.849\textsuperscript{\emph{e}} & 
			$-40.556$\textsuperscript{\emph{f}} \\
			1,000 / 1,437 & 30.3 & 8.8 & 0.3 & 0.1 & 28.755 & 1.010 & 1.183 & 22.107 & 8.467 
			& $-1.934$ \\
			5,000 / 5,793 & 39.4 & 21.4 & 1.8 & 0.2 & 9.919 & 0.337 & 0.354 & 9.546 & 4.161 
			& 1.507 \\
			10,000 / 11,603 & 54.5 & 28.6 & 3.5 & 0.4 & 5.258 & 0.157 & 0.161 & 5.093 & 
			1.687 & 0.699 \\
			50,000 / 92,707 & 87.9 & 70.8 & 14.9 & 2.5 & 0.906 & 0.031 & 0.031 & 0.902 & 
			0.341 & 0.358 \\
			100,000 / 184,903 & 90.9 & 78.5 & 22.7 & 4.4 & 0.483 & 0.011 & 0.011 & 0.483 & 
			0.177 & 0.228 \\
			500,000 / 703,445 & 97.0 & 89.3 & 46.0 & 13.5 & 0.071 & $-0.001$ & $-0.001$ & 0.071 
			& 0.049 & 0.050 \\
			1,000,000 / 1,215,321 & 97.0 & 92.8 & 56.3 & 19.5 & 0.029 & $-0.001$ & $-0.001$ &
			0.029 & 0.017 & 0.017 \\
			\hline\hline
		\end{tabular}
\vspace*{1mm}

	 {\footnotesize
	 	\textsuperscript{\emph{a}} The equilibrium geometry, $R = R_{e}$, and 
	 	the geometries that represent a simultaneous stretching of both O--H bonds 
	 	by factors of 2 and 3 without changing the $\angle\text{(H--O--H)}$ angle 
	 	were taken from ref \citenum{olsen-h2o}.
	 	\textsuperscript{\emph{b}} \%S, \%D, \%T, and \%Q are, respectively, the 
	 	percentages of the singly, doubly, triply, and quadruply excited $S_z = 0$ 
	 	determinants of $A_1$ symmetry captured during the CIPSI computations.
	 	\textsuperscript{\emph{c}} Errors relative to FCI in millihartree (see Table 
	 	\ref{table1} for the FCI energies).
		\textsuperscript{\emph{d}} Equivalent to RHF.
		\textsuperscript{\emph{e}} Equivalent to CCSD.
		\textsuperscript{\emph{f}} Equivalent to CR-CC(2,3).
	}
\end{table*}
\end{landscape}

\pagebreak

\begin{landscape}
\begin{table*}
	\caption{\label{table3} Convergence of the energies resulting from the 
	all-electron CIPSI 
	calculations initiated from the CISD wave function and the corresponding 
	CIPSI-based ec-CC energies toward FCI for the H$_2$O molecule, as described by 
	the cc-pVDZ basis set,\cite{Dunning1989} at $R = 2 R_{e}$.\textsuperscript{\emph{a}}}
\begin{tabular}{c @{\extracolsep{18pt}} r @{\extracolsep{16pt}} r @{\extracolsep{16pt}}
r @{\extracolsep{16pt}} r @{\extracolsep{18pt}} r @{\extracolsep{4pt}} r @{\extracolsep{4pt}} r @{\extracolsep{18pt}}
r @{\extracolsep{10pt}} r @{\extracolsep{10pt}}r}
		\hline\hline
                \multirow{2}{*}{$N_\text{det(in)}$ / $N_\text{det(out)}$} &
		\multirow{2}{*}{\%S\textsuperscript{\emph{b}}} & 
		\multirow{2}{*}{\%D\textsuperscript{\emph{b}}} & 
		\multirow{2}{*}{\%T\textsuperscript{\emph{b}}} & 
		\multirow{2}{*}{\%Q\textsuperscript{\emph{b}}} &
		\multicolumn{3}{c}{CIPSI\textsuperscript{\emph{c}}} & 
		\multicolumn{3}{c}{ec-CC\textsuperscript{\emph{c}}} \\
		\cline{6-8} \cline{9-11}
		& & & & & \multicolumn{1}{c}{$E_\text{var}$} & 
		$E_\text{var} + \Delta E^{(2)}$ &
                $E_\text{var} + \Delta E_{r}^{(2)}$ &
		\multicolumn{1}{c}{I} & \multicolumn{1}{c}{II} & 
		\multicolumn{1}{c}{II$_3$} \\
		\hline
		5,000  / 6,887  & 100 & 100 & 1.5 & 0.2 & 20.034 & 1.755 & 1.829 & 20.034 & 9.408 & 
		3.390 \\
		10,000 / 13,788 & 100 & 100 & 4.3 & 0.5 & 10.053 & 0.464 & 0.481 & 10.053 & 3.085 & 
		0.374 \\
		50,000 / 55,194 & 100 & 100 & 14.8 & 2.2 & 3.422 & 0.125 & 0.127 & 3.422 & 0.920 & 
		0.434 \\
		100,000 / 110,334 & 100 & 100 & 23.6 & 4.5 & 2.472 & 0.076 & 0.077 & 2.472 & 0.638 & 
		0.410 \\
		500,000 / 793,987 & 100 & 100 & 61.5 & 22.9 & 0.239 & 0.009 & 0.009 & 0.239 & 0.117 
		& 0.109 \\
		1,000,000 / 1,476,373 & 100 & 100 & 74.1 & 33.1 & 0.119 & 0.004 & 0.004 & 0.119 & 
		0.063 & 0.062 \\
			\hline\hline
		\end{tabular}
\vspace*{1mm}

{\footnotesize
	\textsuperscript{\emph{a}} The geometry representing a simultaneous stretching 
	of both O--H bonds 
	by a factor of 2 without changing the $\angle\text{(H--O--H)}$ angle 
	was taken from ref \citenum{olsen-h2o}. 
	For this problem, the CISD wave function contains 3,416 $S_z = 0$ determinants 
	of $A_1$ symmetry.
	\textsuperscript{\emph{b}} \%S, \%D, \%T, and \%Q are, respectively, the 
	percentages of the singly, doubly, triply, and quadruply excited $S_z = 0$ 
	determinants of $A_1$ symmetry captured during the CIPSI computations.
	\textsuperscript{\emph{c}} Errors relative to FCI in millihartree (for the FCI 
	energy at $R = 2R_{e}$, see Table \ref{table1}).
}
\end{table*}
\end{landscape}

\begin{table*}
\caption{\label{table4}
The three categories of the ec-CC models analyzed and discussed in this article.
}
\begin{tabular}{l p{5in}}
\hline\hline
ec-CC model & meaning \\
\hline
ec-CC-I &
the ec-CC approach in which one solves eqs \ref{fcc-singles} and \ref{fcc-doubles} for
$T_{1}$ and $T_{2}$ in the presence of $T_{3}$ and $T_{4}$ extracted from CI without making any
{\it a posteriori} modifications in the $T_{3}$ and $T_{4}$ operators \\
ec-CC-II &
the ec-CC approach in which one solves eqs \ref{fcc-singles} and \ref{fcc-doubles} for
$T_{1}$ and $T_{2}$ in the presence of $T_{3}$ and $T_{4}$ extracted from CI after eliminating
the $T_{3}$ and $T_{4}$ components that do not have the companion $C_{3}$ and $C_{4}$ amplitudes \\
ec-CC-II$_{3}$ &
the ec-CC approach obtained by correcting the ec-CC-II energy for the $T_{3}$ correlations
that do not have the companion $C_{3}$ coefficients in CI, omitted in ec-CC-II calculations \\
\hline\hline
\end{tabular}
\end{table*}


\clearpage

\newpage

\begin{chart}
\centering
\includegraphics[scale=1.2]{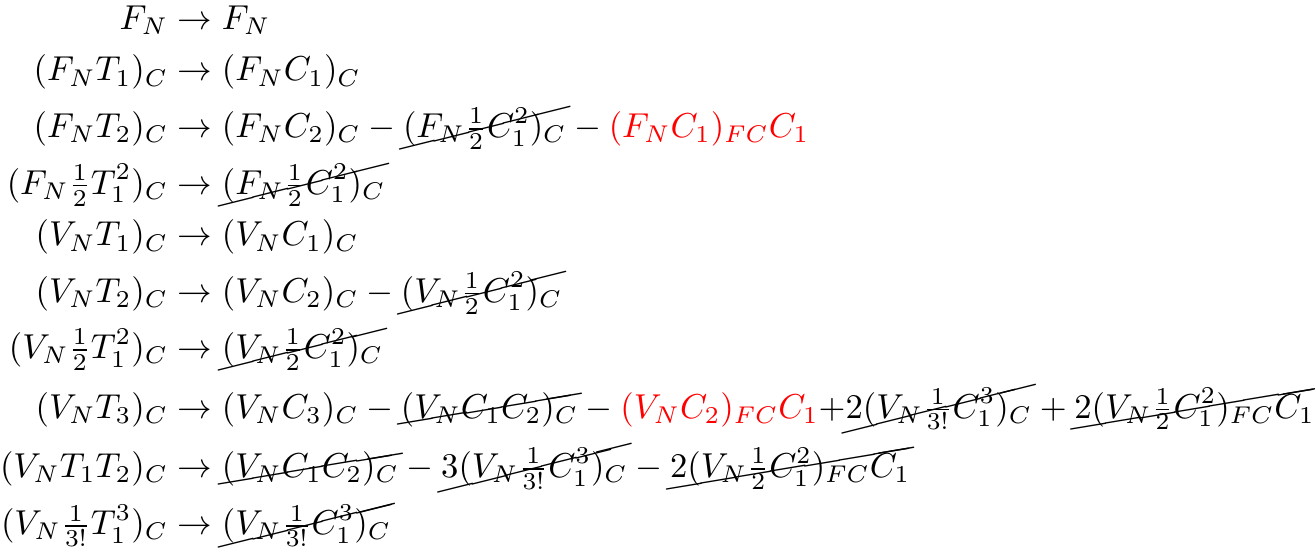}
\vspace*{12pt}
\caption{\label{chart-B1}
The correspondence between the various $(H_N e^T)_C$ contributions to
the CC equations projected on the singly excited determinants, eq \ref{fcc-singles},
and their counterparts resulting from the application of eq \ref{cluster_analysis}.
The unlinked terms in which the fully connected operator products
$(F_N \mbf{C_1})_{FC}$ and $(V_N \mbf{C_2})_{FC}$ that contribute to the correlation
energy multiply $C_1$, resulting in eq \ref{B2}, are highlighted in red.
}
\end{chart}

\newpage

\begin{chart}
\centering
\includegraphics[scale=1.2]{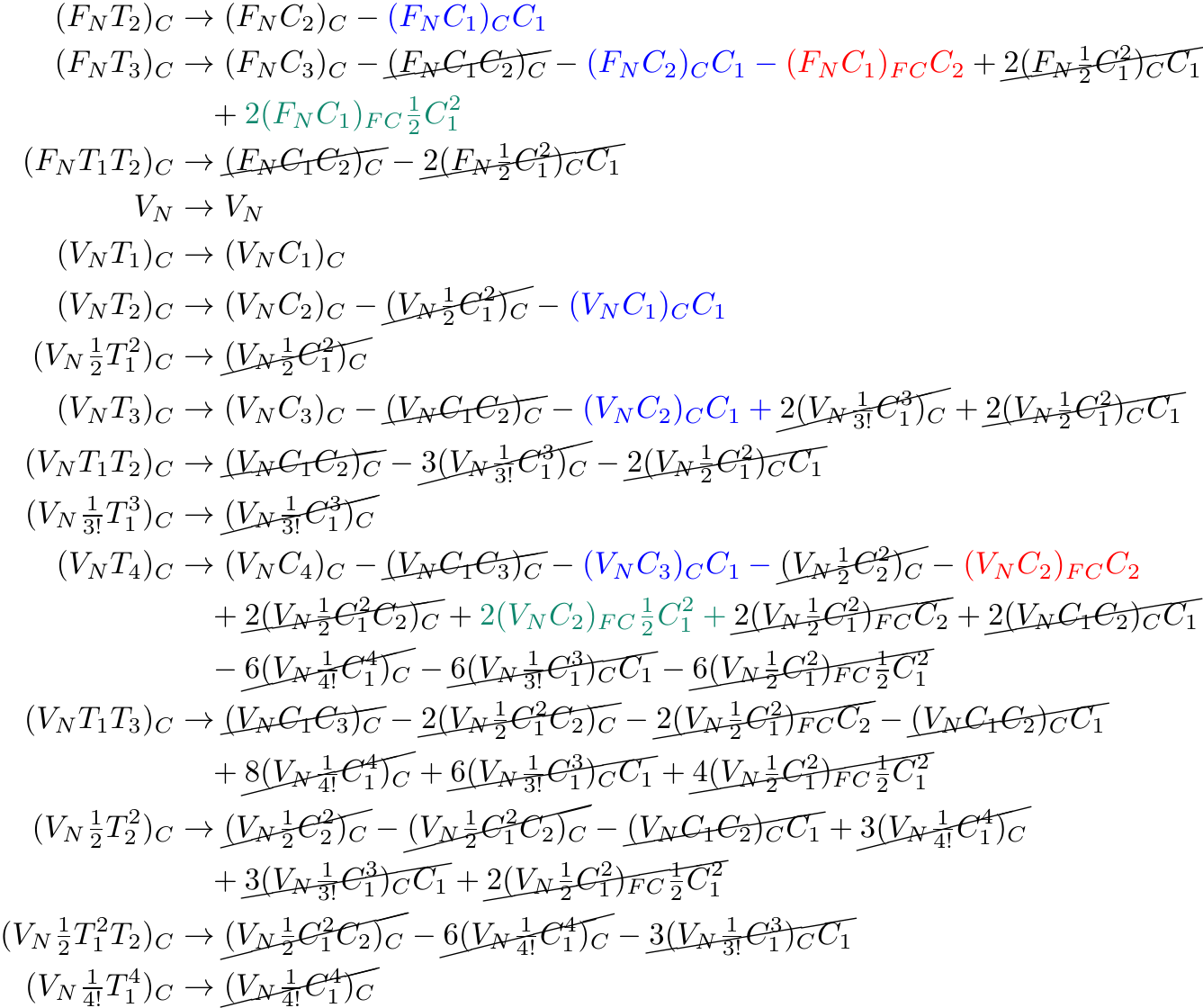}
\vspace*{12pt}
\caption{\label{chart-B2}
The correspondence between the various $(H_N e^T)_C$ contributions to
the CC equations projected on the doubly excited determinants, eq \ref{fcc-doubles},
and their counterparts resulting from the application of eq \ref{cluster_analysis}.
The unlinked terms in which the fully connected operator products $(F_N \mbf{C_1})_{FC}$ and $(V_N \mbf{C_2})_{FC}$
that contribute to the correlation energy multiply $C_2$, resulting in eq \ref{B9}, are highlighted in red.
The unlinked terms in which the fully connected operator products $(F_N \mbf{C_1})_{FC}$ and $(V_N \mbf{C_2})_{FC}$
that contribute to the correlation energy multiply $\tfrac{1}{2} C_1^2$, resulting in eq \ref{B8}, are
highlighted in green.
The disconnected terms in which the $(F_N \mbf{C_1})_C$,
$(F_N \mbf{C_2})_C$, $(V_N \mbf{C_1})_C$, $(V_N \mbf{C_2})_C$, and $(V_N \mbf{C_3})_C$
connected operator products multiply $C_1$, resulting in eq \ref{B7}, are highlighted in blue.
}
\end{chart}

\newpage

\end{document}